\documentclass[]{aa} 
\usepackage{graphicx}
\usepackage{lscape}
\usepackage{longtable}
\usepackage{arydshln}
\usepackage{multirow}
\usepackage{amsmath}
\usepackage{enumitem}
\usepackage{epstopdf}
\usepackage{natbib}
\usepackage[htt]{hyphenat}

\newcommand{\hhco}{H$_2$CO}
\newcommand{\chhhoh}{CH$_3$OH}
\newcommand{\chhco}{H$_2$CCO}
\newcommand{\chhhcho}{CH$_3$CHO}
\newcommand{\chhhcn}{CH$_3$CN}
\newcommand{\chhhnc}{CH$_3$NC}
\newcommand{\chhhcch}{CH$_3$CCH}
\newcommand{\cch}{C$_2$H}
\newcommand{\ccchh}{C$_3$H$_2$}
\newcommand{\ccch}{C$_3$H}

\bibpunct{(}{)}{;}{a}{}{,} 

\begin{document}

\title{Abundances of sulphur molecules in the Horsehead nebula} 
\subtitle{First NS$^+$ detection in a photodissociation region}
   \author{P. Rivi\`ere-Marichalar\inst{1,2}, A. Fuente\inst{2}, J. R. Goicoechea\inst{1}, J. Pety\inst{3, 4}, R. Le Gal\inst{5}, P. Gratier\inst{6}, V. Guzm\'an\inst{7}, E. Roueff\inst{8}, J. C. Loison\inst{9}, V. Wakelam\inst{6}, M. Gerin\inst{4}}

   \institute{Instituto de F\'{\i}sica Fundamental (CSIC), Calle Serrano 121, 28006 Madrid, Spain 
                   \email{p.riviere@oan.es}
                   \and Observatorio Astron\'omico Nacional (OAN,IGN), Apdo 112, E-28803 Alcal\'a de Henares, Spain  
                   \and Institut de Radioastronomie Millim\'etrique (IRAM), 300 rue de la Piscine, 38406 Saint Martin d'H\`eres, France 
                   \and LERMA, Observatoire de Paris, PSL Research University, CNRS, Sorbonne Universit\'es, UPMC Univ. Paris 06, Ecole Normale
                           Sup\'erieure, F-75005 Paris, France  
                   \and Harvard-Smithsonian Center for Astrophysics, 60 Garden St., Cambridge, MA 02138, USA 
                   \and Laboratoire d'Astrophysique de Bordeaux, Univ. Bordeaux, CNRS, B18N, all\'ee Geoffroy Saint-Hilaire, 33615 Pessac, France 
                   \and Instituto de Astrof\'isica, Pontificia Universidad Cat\'olica de Chile, Av. Vicu\~na Mackenna, 4860, 7820436, Macul, Santiago, Chile 
                   \and LERMA, Observatoire de Paris, PSL Research University, CNRS, Sorbonne Universit\'es, UPMC Univ. Paris 06, F-92190 Meudon, France 
                   \and Institut des Sciences Mol\'eculaires de Bordeaux (ISM), CNRS, Univ. Bordeaux, 351 cours de la Lib\'eration, 33400, Talence, France 
   }
   \authorrunning{Rivi\`ere-Marichalar et al.}
   \date{}

 \abstract 
{Sulphur is one of the most abundant elements in the Universe (S/H$\sim$1.3$\times$10$^{-5}$) and plays a crucial role in biological systems on Earth. The understanding of its chemistry is therefore of major importance. }
{Our goal is to complete the inventory of S-bearing molecules and their abundances in the prototypical photodissociation region (PDR) the Horsehead nebula to gain insight into sulphur chemistry in UV irradiated regions. Based on the WHISPER (Wide-band High-resolution Iram-30m Surveys at two positions with Emir Receivers) millimeter (mm) line survey, our goal is to provide an improved and more accurate description of sulphur species and their abundances towards the core and PDR positions in the Horsehead
}
{The Monte Carlo Markov Chain (MCMC) methodology and the molecular excitation and radiative transfer code RADEX were used to explore the parameter space and determine physical conditions and beam-averaged molecular abundances.
}
{A total of 13 S-bearing species (CS, SO, SO$\rm _{2}$, OCS, H$\rm _{2}$CS -- both ortho and para -- HDCS,
C$\rm _{2}$S, HCS$\rm ^{+}$, SO$\rm ^{+}$, H$_2$S, S$_2$H, NS and NS$^+$) have been detected in the two targeted positions. This is the first detection of SO$^+$ in the Horsehead and the first detection of NS$^+$ in any PDR. We find a differentiated chemical behaviour between C-S and O-S bearing species within the nebula. The C-S bearing species C$_2$S and o-H$\rm _{2}$CS present fractional abundances a factor of $>$ two  higher  in the core than in the PDR.  In contrast, the O-S bearing molecules  SO, SO$_2,$ and OCS present similar abundances towards both positions. A few molecules, SO$^+$, NS, and NS$^+$,  are more abundant towards the PDR than towards the core, and could be considered as PDR tracers.}
{This is the first complete study of S-bearing species towards a PDR. Our study shows that CS, SO, and H$_2$S
are the most abundant S-bearing molecules in the PDR with abundances of $\sim$ a few 10$^{-9}$. We recall that SH, SH$^+$, S, and S$^+$ are not observable at the wavelengths covered by the WHISPER survey. At the spatial scale of our observations, the total abundance of S atoms locked in the detected species is $<$ 10$^{-8}$, only $\sim$0.1\% of the cosmic sulphur abundance.}


\maketitle

\begin{figure*}[!t]
\begin{center}
 \includegraphics[scale=1.2]{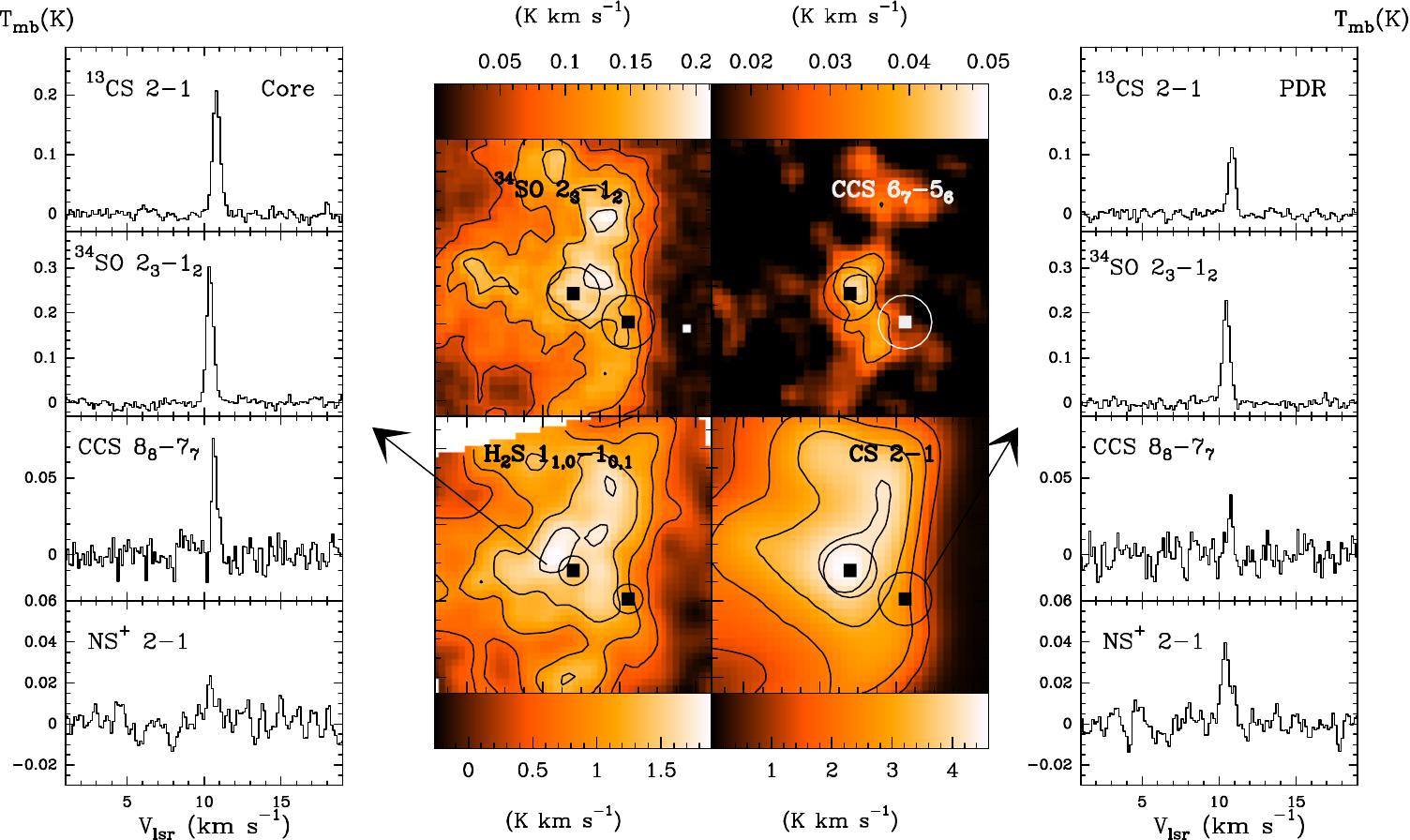}\\
 \caption{Integrated intensity maps of the
$^{13}$CS 2$\rightarrow$1, $^{34}$SO 2$_3$$\rightarrow$1$_2$, CCS 8$_7$$\rightarrow$7$_6,$ 
and H$_2$S 1$_{1,0}$$\rightarrow$1$_{0,1}$ maps as observed with the 30m IRAM telescope. The beam of the 30m telescope at 3mm (HPBW$\sim$29$"$) is indicated by two circles centred on the positions targeted within the  Whisper spectral survey.  Line contours depict 30 to 90 by 15\% levels with respect to the emission peak.}
\label{Fig:Map}
\end{center}
\end{figure*}

\section{Introduction} 
Sulphur is one of the most abundant elements in the Universe \citep[S/H$\sim$1.3$\times$10$^{-5}$ in the solar photosphere, ][]{Asplund2005} and plays a crucial role in biological systems on Earth. Surprisingly, sulphuretted molecules are not as abundant as expected in the interstellar medium. A few sulphur compounds have been detected in diffuse clouds demonstrating that the sulphur abundance
in these low density regions is close to the cosmic value \citep{Neufeld2015}. A moderate sulphur depletion 
(a factor of four) is estimated in the external layers of the photodissociation region (PDR) in the Horsehead nebula
\citep{Goicoechea2006}. In cold molecular clouds, a large depletion of sulphur is usually considered to reproduce the observations \citep[see for instance][]{Tieftrunk1994, Vastel2018}. The depletion of sulphur is observed not only in cold pre-stellar cores, but also in hot cores/corinos \citep{Wakelam2004}. One would expect that most of the sulphur is 
locked in the icy grain mantles in dense cores but we should see almost all sulphur return to the gas phase
in hot cores and strong shocks. However, even in the well-known Orion$-$KL hot core where the icy grain mantles 
are expected to evaporate releasing the molecules to the gas phase, one needs to assume a sulphur depletion 
of a factor of approximately ten  to reproduce the observations \citep{Esplugues2014, Crockett2014}. Because of the high hydrogen abundances and the mobility of hydrogen in the ice matrix, sulphur atoms impinging on interstellar ice mantles are expected to form H$_2$S preferentially. However, there are only upper limits of the solid H$_2$S abundance (e.g. \citealp{JimenezEscobar2011}). \cite{Oba2018} show that, at high densities, chemical desorption is likely to exceed photodesorption, thus becoming an efficient way of removing H$_2$S from the surface of grains.

Thus far, OCS is the only S-bearing molecule unambiguously detected in interstellar ice because of its large band strength in the infrared \citep{Geballe1985,Palumbo1995}; tentatively, SO$_2$ has also been detected \citep{Boogert1997}. The bulk of the S-budget might be locked in atomic sulphur in the gas phase and/or in S-polymers in the solid phase, which are not easily observable \citep{JimenezEscobar2014,Wakelam2004}. The recent detection of the doubly-sulphuretted species S$_2$H in the Horsehead nebula adds important  information as well as new questions to the overall complex sulphur chemistry problem, possibly supporting the scenario of an important role of surface chemistry in the formation of the gaseous sulphuretted compounds \citep{Fuente2017}. Observations of the comet 67P with Rosetta showed that H$_2$S and atomic S are the most important S-bearing species in cometary ices, with an abundance of 0.015~\% relative to water \citep{Calmonte2016, Bockelee2017}. 

Only a few PDRs have been investigated in detail taking advantage of their proximity and favourable edge-on geometry: NGC 7023 \citep{Fuente1993, Fuente1996b, Fuente2000}, the Orion Bar \citep{Cuadrado2015, Cuadrado2017, Goicoechea2016,Goicoechea2017},  Mon R2 \citep{Ginard2012,Trevino2014,Trevino2016,Pilleri2012,Pilleri2013,Pilleri2014}, and the Horsehead nebula. Here we investigate the sulphur chemistry in the Horsehead nebula. At a distance of 400 pc, the Horsehead is a PDR viewed nearly edge-on and illuminated  by  the  O9.5V star $\sigma$Ori at a  projected  distance of $\sim$3.5 pc. The intensity of the incident far ultra-violet (FUV) radiation field is $\chi$=60  \citep{Pety2005} relative to the interstellar radiation field in Draine units \citep{Draine1978}. This PDR presents a differentiated chemistry from others associated with warmer regions such as the Orion Bar, NGC 7023, and Mon R2. One main difference is that the dust temperature is around $\sim$20-30 K in the PDR \citep{Goicoechea2009}, meaning it is below or close to the sublimation temperature of many species, allowing a rich surface chemistry on the irradiated surfaces as has been demonstrated by recent observational and theoretical studies \citep{Guzman2011,Guzman2013}. A first analysis of the sulphur chemistry was carried out by Goicoechea et al. (2006) who derived a gas-phase sulphur abundance of S/H=(3.5$\pm$1.5)$\times$10$^{-6}$ in the low extinction PDR on the basis of single-dish and interferometric observations of the CS and HCS$^+$ millimeter lines.  

The Horsehead Wide-band High-resolution Iram-30m Surveys at two positions with Emir Receivers (WHISPER) project is a complete unbiased line survey of the 3, 2, and 1 mm bands using the Institut de Radioastronomie Millim\'etrique (IRAM) 30m telescope. WHISPER has provided valuable hints regarding the chemistry of this region. The detection of the molecular ion CF$^+$ towards the HCO peak is well understood in terms of gas-phase photochemistry \citep{Guzman2012}. Photo-destruction of large polyatomic molecules or small grains into smaller hydrocarbon precursors  is invoked to explain the extraordinarily high abundance of small hydrocarbons, such as {\cch}, {\ccchh}, {\ccch,}~and C$_3$H$^+$~\citep{Pety2012,Guzman2015}. The detection of several complex organic molecules (COMs) towards the warm ($T_\mathrm{kin}\sim60$ K) PDR was unexpected. In fact, the chemical complexity reached in the Horsehead is surprisingly high, with COMs of up to seven atoms: HCOOH, \chhco, \chhhcho,~and {\chhhcch}~\citep{Guzman2014}. Current pure gas-phase models cannot reproduce the inferred {\hhco}, {\chhhoh,} and COMs abundances in the Horsehead PDR~\citep{Guzman2011,Guzman2013}, which supports the theory of a grain surface origin for these molecules. \citet{LeGal2017} were able to reproduce the observed COMs abundances using a chemical model with grain surface chemistry and proposed that chemical desorption, instead of photodesorption, was most likely the dominant process to release COMs to the gas phase. However, the efficiency of chemical desorption seems to be extremely dependent on the species involved and the chemical composition of the icy mantle \citep{Minissale2016,Oba2018}. {\chhhcn} and {\chhhnc}, key species for the formation of prebiotic molecules, seem to have a very specific formation pathway in this PDR~\citep{Gratier2013}. In this paper we use the data of the Horsehead WHISPER project to investigate the chemistry of  S-bearing species in this prototypical nebula. 

\section{Observations and data reduction}\label{sec:Sample}
The data used in this work are taken from the Horsehead WHISPER  (Wide-band High-resolution Iram-30m Surveys at two positions with Emir Receivers, PI: J. Pety) project. The Horsehead WHISPER project is a complete unbiased line survey of the 3, 2, and 1 mm bands using the IRAM 30m telescope. Two positions were observed within this survey: i) the PDR position, where HCO emission peaks (RA=5$^{\rm h}$40$^{\rm m}$53$^{\rm s}$.936, Dec=02$^\circ$28$'$00$''$, J2000) at the UV-illuminated surface of the Horsehead nebula \citep[][hereafter, PDR]{Gerin2009}, and ii) the core, where DCO$^+$ peaks (RA=5$^{\rm h}$40$^{\rm m}$55$^{\rm s}$.61, Dec=02$^\circ$27$'$38$''$, J2000), which corresponds to a cold \citep[$T_{k}\sim10 -20~K$, ][]{Pety2007} and UV-shielded condensation located less than 40$''$ away from the PDR edge (hereafter, core). During the observations we used the position-switching procedure with the reference position located at an offset ($-$100$''$,0) relative to RA: 05$^{\rm h}$40$^{\rm m}$54$^{\rm s}$.27 Dec: $-$02$^\circ$28$'$00$''$.0. In order to avoid relative calibration errors we integrated towards the HCO peak and the DCO$^+$ peak  alternatively, changing from one position to the other every 15 minutes. Line intensities are given in main brightness temperature (T$\rm _{MB}$) and the frequency resolution in our survey is 49 kHz. The uncertainties in T$\rm _{MB}$ due to calibration errors are $\sim$10\%. 

In Fig. \ref{Fig:Map}, we show the integrated intensity maps of the $^{34}$SO 2$_3$$\rightarrow$1$_2$, CCS 8$_7$$\rightarrow$7$_6$ H$_2$S 1$_{1,0}$$\rightarrow$1$_{0,1}$, and  CS 2$\rightarrow$1 lines as observed with the IRAM 30m telescope. These maps were observed in previous projects by our team. The two targeted positions within WHISPER, PDR and core, are marked with filled squares and the 3mm and 2mm beam of the 30m telescope are also shown. The H$_2$S 1$_{1,0}$$\rightarrow$1$_{0,1}$ line has been presented in \cite{Fuente2017}. The other lines were observed in November and December 2017 under good winter weather conditions (2 mm of precipitable water vapour) using the Eight MIxer Receiver (EMIR) sideband-separation receivers. The position-switching, on-the-fly observing mode was used, with off-position offsets ($\delta$RA, $\delta$Dec) = $(-100'',0'')$, that is, into the HII region. We observed along and perpendicular to the direction of the exciting star in zigzags, covering an area of $200''\times200''$. The maps were processed with the Grenoble Image and Line Data Analysis System (GILDAS\footnote{See  \texttt{http://www.iram.fr/IRAMFR/GILDAS} for more information about the GILDAS softwares.}) software \citep{Pety2005}.  The data were first calibrated to the $\rm T_{a}^{*}$ scale using the chopper-wheel method~\citep{Penzias1973}, and then converted into main-beam temperatures ($\rm T_{mb}$) using the forward and main-beam efficiencies of the telescope. The resulting spectra were then baseline-corrected and gridded through convolution with a Gaussian to obtain the maps. Beam sizes are roughly in the range 12$\arcsec$ to 25$\arcsec$, depending on the exact frequency of each transition, more precisely, HPBW($\rm \arcsec$) = 2460/$\rm \nu$(GHz)\footnote{See \texttt{http://www.iram.es/IRAMES/telescope/telescope \linebreak Summary/telescope\_summary.html.}}.

\section{Molecular inventory}\label{Sec:results}

Taking the high sensitivity WHISPER survey as the observational basis, we have systematically searched for all the S-bearing species included in the Cologne Database for Molecular Spectroscopy (CDMS) catalogue. A complete list of intensities for the detected lines is provided in Table~\ref{Tab:lineFluxes}. Figure~\ref{Fig:allDetections} depicts the spectra of the detected transitions grouped by species. 
For the detected species, we have carried out a multi-transitional study that includes lines of different isotopologues, to determine accurate molecular abundances (Tables~\ref{Tab:rotational_diagrams}, and  \ref{Tab:RADEX_CD}). In the case of non-detections, an upper limit to the fractional abundance is given. The list of non-detected species is shown in Table~\ref{Tab:CD_UL}. A detailed description of the methods used to derive molecular abundances can be found in Sects. \ref{Sec:rot_diag} and \ref{Sec:RADEX_models}. 

\begin{figure}[]
\begin{center}
 \includegraphics[scale=0.25,trim = 0mm 0mm 0mm 0mm,clip]{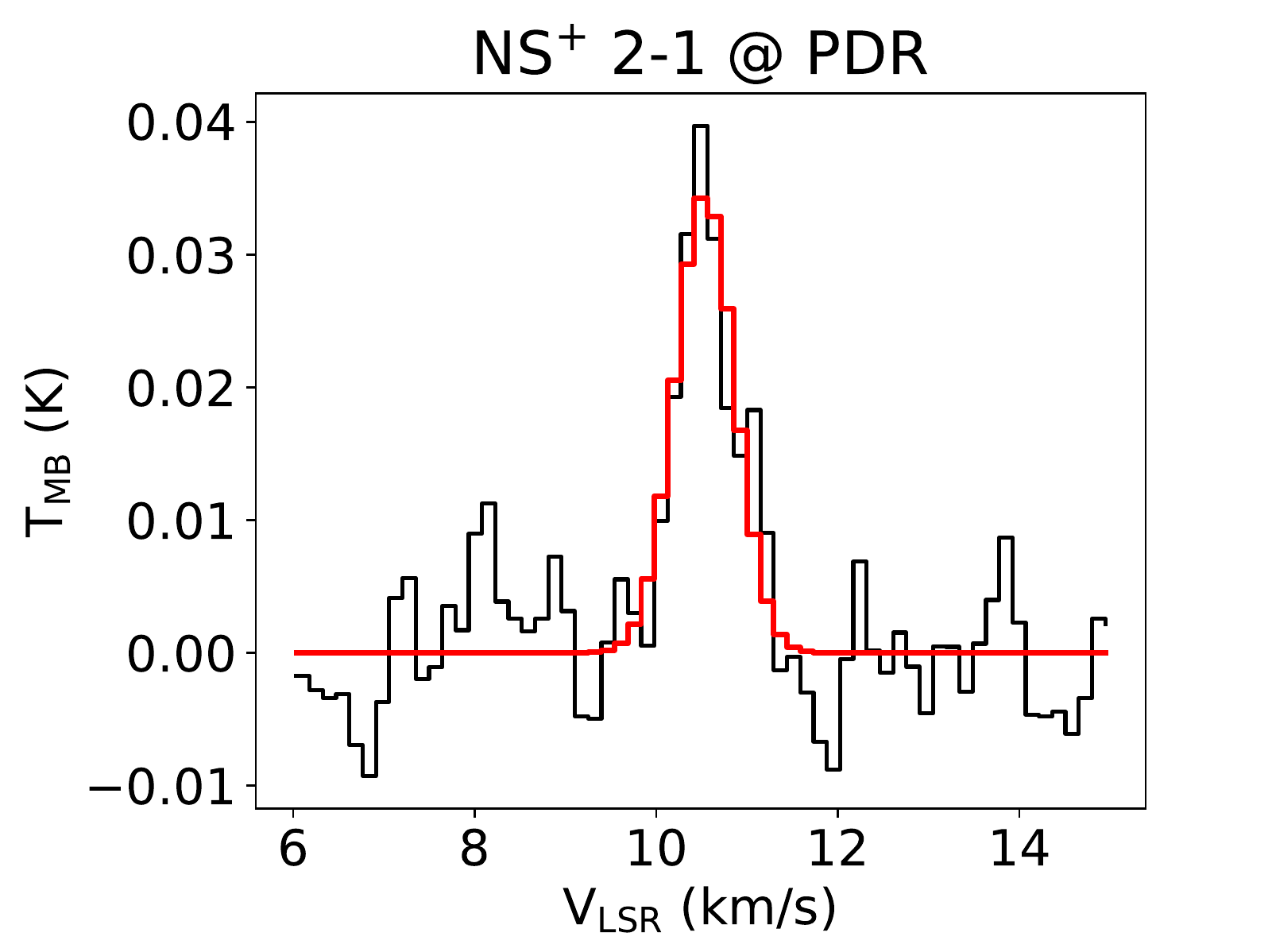}
 \includegraphics[scale=0.25,trim = 0mm 0mm 0mm 0mm,clip]{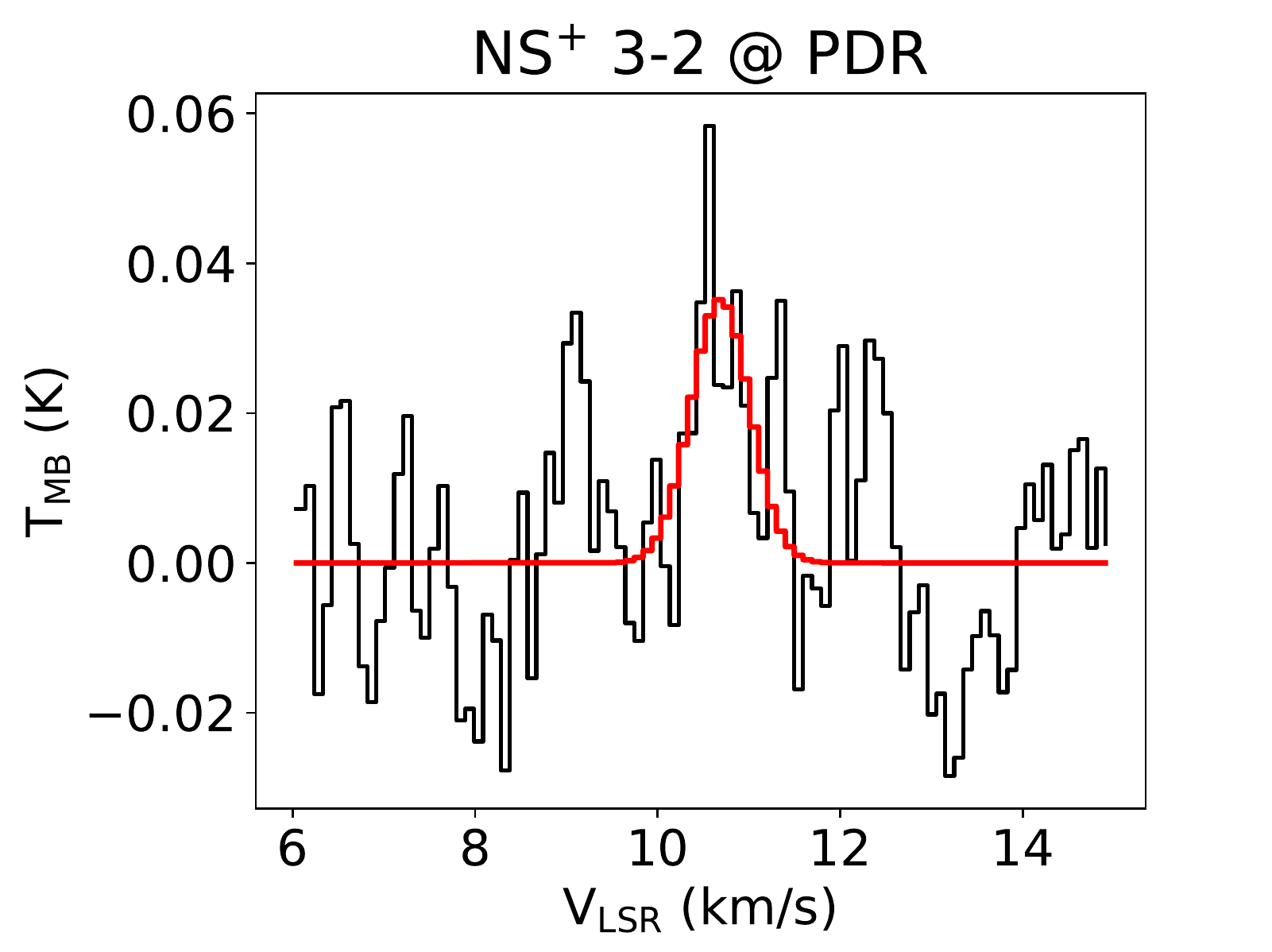}\\
 \includegraphics[scale=0.25,trim = 0mm 0mm 0mm 0mm,clip]{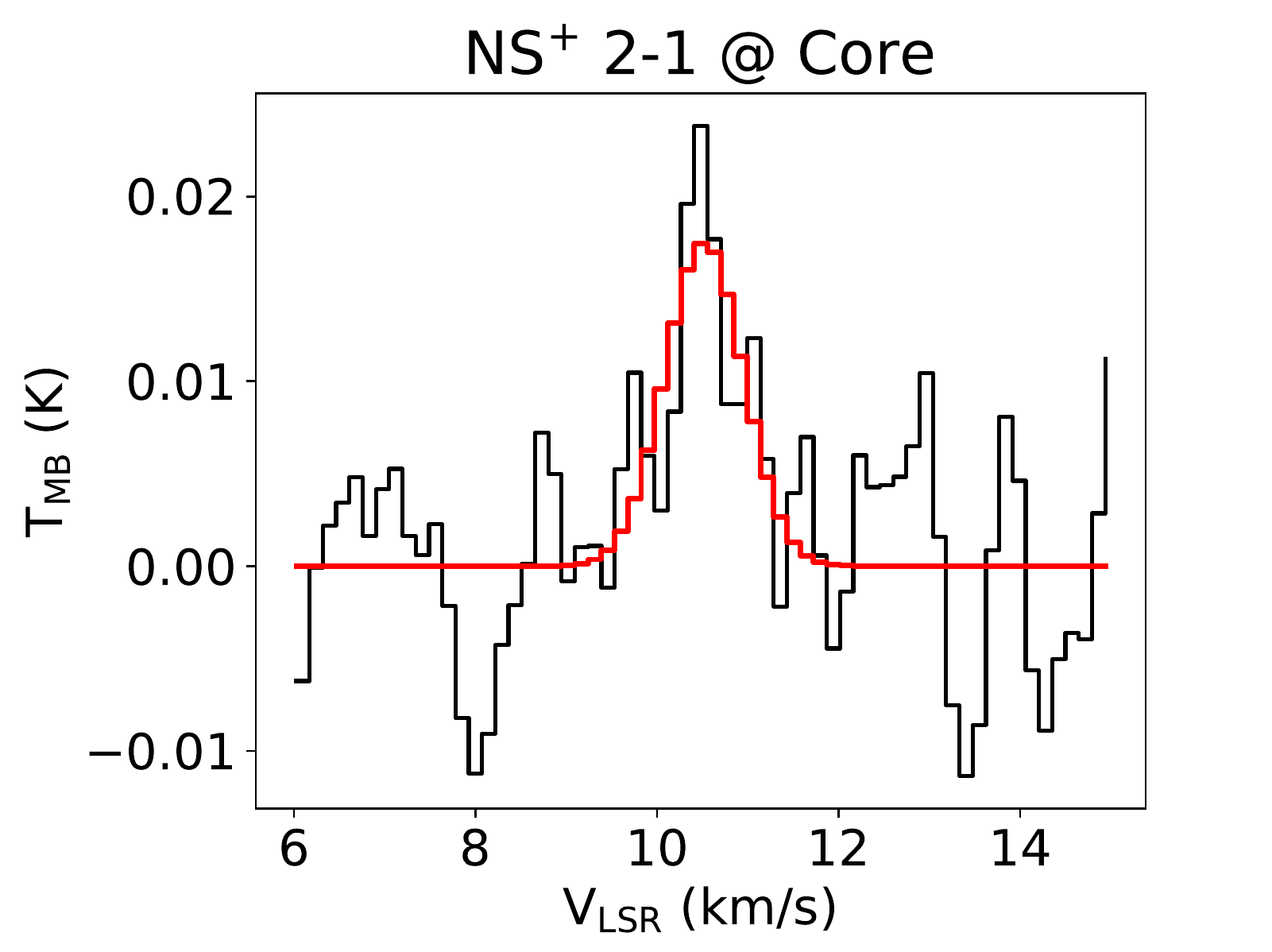} 
 \includegraphics[scale=0.25,trim = 0mm 0mm 0mm 0mm,clip]{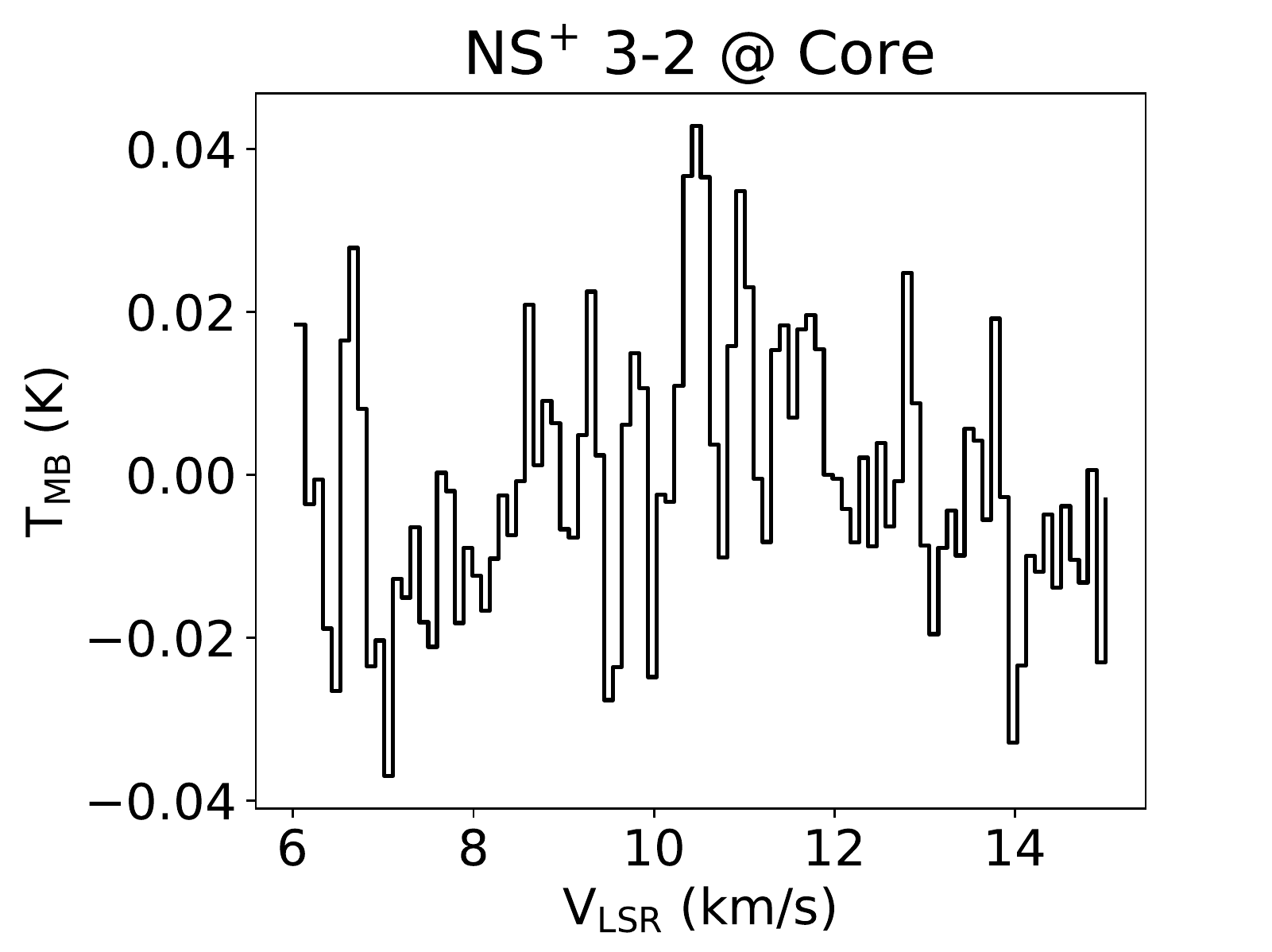}  
 \caption{Spectra of NS$\rm ^{+}$  in the PDR and the core. The red line shows the Gaussian fit to the observed line profile.}
 \label{Fig:NSplus_spec}
\end{center}
\end{figure}

The list of S-bearing species detected in the Horsehead includes CS, SO, SO$\rm _{2}$, OCS, H$\rm _{2}$CS (both ortho- and para-),  HDCS, C$\rm _{2}$S, HCS$\rm ^{+}$, SO$\rm ^{+}$, H$_2$S, S$_2$H, NS, and NS$^+$. The detections of H$_2$S and S$_2$H were previously reported by \citet{Fuente2017}. It is important to note that this list constitutes the most complete inventory of S-bearing species in a PDR thus far.  We have not detected DCS$^+$ and CCCS towards any of the observed positions, and we provide upper limits for the fractional abundance of these molecules. The detection of NS$^+$ reported here is worth noticing as the first detection of NS$^+$ in a PDR. NS$^+$ was recently identified by  \citet{Cernicharo2018} in 
a variety of dark clouds, with a NS/$\rm NS^{+}$$\sim$ 30$-$50. However, a  NS/$\rm NS^{+}$$>$ 3000 was measured in the Orion KL hot core \citep{Cernicharo2018}. We have detected two NS$^+$ lines, J=2$\rightarrow$1 at  110198.5 MHz and J=3$\rightarrow$2 at 150295.6 MHz, with a signal-to-noise ratio (S/N) $>$4 towards the PDR position. Only the line at 110198.5 MHz has been detected towards the core (see Fig. \ref{Fig:NSplus_spec}). 

We present also the first detection of SO$\rm ^+$ in a low-UV PDR ($\chi$$<$500). Detected earlier by \citet{Turner1992}, SO$^+$ has been 
detected in variety of regions, which include massive star formation regions, shocked regions, and translucent clouds \citep{Turner1992,Turner1994,Turner1996}.
Although widely detected, a high abundance of SO$^+$ relative to SO, X(SO$^+$)/X(SO)$>$100, seems to be related to photon-dominated chemistry and was proposed by \citet{Sternberg1995} as a tracer of UV illuminated regions. These high SO$^+$  abundances have been previously measured in the PDRs associated to NGC 7023 \citep{Fuente2003}, the Orion Bar \citep{Fuente2003, Goicoechea2017}, and  
Monoceros R 2 \citep{Ginard2012}, but this ion has not been detected in a low-UV  PDR such as the Horsehead thus far. 

\begin{table*}[]
\caption{Rotational temperatures, column densities,  and abundances derived from rotational diagrams for S-bearing species in the Horsehead nebula.}             
\label{Tab:rotational_diagrams}      
\centering          
\begin{tabular}{l | lcc | lcc}     
\hline \hline 
              & \multicolumn{3}{c}{PDR} &  \multicolumn{3}{|c}{Core} \\
\hline              
Species & $\rm T_{rot}$ & N(X) & Abundance & $\rm T_{rot}$ & N(X) & Abundance \\
             & (K) & ($\rm cm^{-2}$) & $\rm {N(X)} \over {N(H)+2N(H_{2})} $ & (K) & ($\rm cm^{-2}$) & $\rm {N(X)} \over {N(H)+2N(H_{2})} $ \\
\hline
CS$\rm ^{*}$     & 6.3    & (3.0$\rm \pm$0.6)$\rm \times 10^{13}$  & (8.0$\rm \pm$2.0)$\rm \times 10^{-10}$ & 5.9 & (5.3$\rm \pm$1.0)$\rm \times 10^{13}$ & (9.0$\rm \pm$2.0)$\rm \times 10^{-10}$ \\
SO$\rm ^{*}$ & 11.6 & (3.2$\rm \pm$0.3)$\rm \times 10^{13}$  & (8.4$\rm \pm$0.8)$\rm \times 10^{-10}$  & 8.2 & (5.0$\rm \pm$0.5)$\rm \times 10^{13}$  & (8.6$\rm \pm$0.9)$\rm \times 10^{-10}$ \\  
p-$\rm H_{2}CS$ & 4.2 & (5.0$\rm \pm$3.0)$\rm \times 10^{11}$ & (1.3$\rm \pm$0.8)$\rm \times 10^{-11}$ & 8.6 & (9.0$\rm \pm$1.0)$\rm \times 10^{11}$ & (1.6$\rm \pm$0.2)$\rm \times 10^{-11}$ \\
o-$\rm H_{2}CS$ & 8.8 & (9.0$\rm \pm$3.0)$\rm \times 10^{11}$ & (2.4$\rm \pm$0.8)$\rm \times 10^{-11}$ & 8.2 & (2.5$\rm \pm$0.8)$\rm \times 10^{12}$ & (4.3$\rm \pm$1.4)$\rm \times 10^{-11}$\\
SO$\rm _{2}$ & 6.5 & (3.0$\rm \pm$0.6)$\rm \times 10^{12}$ & (8.0$\rm \pm$1.6)$\rm \times 10^{-11}$ & 6.9 & (2.9$\rm \pm$0.6)$\rm \times 10^{12}$ & (5.0$\rm \pm$1.0)$\rm \times 10^{-11}$ \\
CCS & 10.3 & (3.0$\rm \pm1.0$)$\rm \times 10^{11}$ & (8.0$\rm \pm$2.6)$\rm \times 10^{-12}$ & 8.1 & (1.5$\rm \pm$0.7)$\rm \times 10^{12}$ & (2.6$\rm \pm$1.2)$\rm \times 10^{-11}$\\         
HCS$\rm ^{+}$ & 11.6 & (5.0$\rm \pm$0.6)$\rm \times 10^{11}$ & (1.3$\rm \pm$0.2)$\rm \times 10^{-11}$ & 10$\rm ^{**}$ & 6$\rm \times 10^{11}$ & 1.0$\rm \times 10^{-11}$ \\
SO$\rm ^{+}$  & 11 & (6.3$\rm \pm$1.8)$\rm \times 10^{11}$ & (1.7$\rm \pm$0.5)$\rm \times 10^{-11}$ & 9 & (4.5$\rm \pm$0.5)$\rm \times 10^{11}$ & (7.8$\rm \pm0.9)$$\rm \times 10^{-12}$\\
HDCS                   &  10$\rm ^{**}$  & (6$\rm \pm$1)$\rm \times 10^{11}$              &   (2$\rm \pm$0.4)$\rm \times 10^{-11}$      &
                               10$\rm ^{**}$  & (1$\rm \pm$0.2)$\rm \times 10^{12}$             &   (2$\rm \pm$0.4)$\rm \times 10^{-11}$     \\
NS                      &  10$\rm ^{**}$  & (4$\rm \pm$0.8)$\rm \times 10^{12}$             &   (1$\rm \pm$0.2)$\rm \times 10^{-10}$      &
                               10$\rm ^{**}$  & (3$\rm \pm$0.6)$\rm \times 10^{12}$             &   (5$\rm \pm$1)$\rm \times 10^{-11}$     \\
NS$^+$              & 10$\rm ^{**}$  & (1$\rm \pm$0.2)$\rm \times 10^{11}$              &   (3$\rm \pm$0.6)$\rm \times 10^{-12}$   & 
                              10$\rm ^{**}$  &  (6$\rm \pm$1)$\rm \times 10^{10}$                &   (1$\rm \pm$0.2)$\rm \times 10^{-12}$       \\
DCS$^+$         &  10$\rm ^{**}$ & $<$1.0$\times$10$^{11}$  &  $<$3$\times$10$^{-12}$   &
                             10$\rm ^{**}$ & $<$1.0$\times$10$^{11}$  &  $<$2$\times$10$^{-12}$  \\
 CCCS            &  10$\rm ^{**}$  & $<$3.2$\times$10$^{11}$   & $<$8$\times$10$^{-12}$ &
                             10$\rm ^{**}$  & $<$3.2$\times$10$^{11}$   & $<$5$\times$10$^{-12}$     \\                            
\hline \hline   
\end{tabular}
\tablefoot{Abundances are computed assuming $\rm N(H_{2}) = 1.9 \times 10^{22}~cm^{-2}$ towards the PDR and $\rm N(H_{2}) = 2.9 \times 10^{22}~cm^{-2}$ towards the core from \cite{Gerin2009}. ($\rm ^{*}$): computed including opacity effects. ($\rm ^{**}$): since only one transition was detected, we assumed $\rm T_{rot}=10~K$ and fitted the column density to match the observed $\rm T_{MB}$ assuming LTE. }    
\end{table*}

\begin{table}
\caption{Collisional partners and collisional excitation rates for the different species modelled with LVG.}
\label{Tab:coll_partners_and_rates}
\centering
\begin{tabular}{ll}
\hline \hline
Species & Collisional excitation rates \\
\hline
CS & \cite{Denis-Alpizar2013} \\
SO & \cite{Lique2006B} \\
SO$\rm _{2}$ & \cite{Green1995} \\
OCS & \cite{Green1978} \\
H$\rm _2$CS & \cite{Wiesenfeld2013}\\
HCS$\rm ^+$ & \cite{Flower1999}$\rm ^*$\\
\hline
\end{tabular}
\tablefoot{ ($\rm ^*$): Adapted from HCO$\rm ^+$ collisional rates.}
\end{table}

\begin{table*}
\caption{Upper limits to the column density and abundance of undetected species.}
\label{Tab:CD_UL}
\begin{center}
\begin{tabular}{lrrrrlll}
\hline
\hline
Molecule    &  \multicolumn{1}{c}{Freq} & \multicolumn{1}{c}{rms$^1$} & \multicolumn{1}{c}{N$_X^2$ }     &  \multicolumn{3}{c}{Abundances wrt H$_2$}  & \multicolumn{1}{c}{Ref}       \\
                  &  \multicolumn{1}{c}{(GHz)} & \multicolumn{1}{c}{(mK)}     & \multicolumn{1}{c}{(cm$^{-2}$)}   &  \multicolumn{1}{c}{Horsehead}  &
\multicolumn{1}{c}{Dark clouds }   & \multicolumn{1}{c}{Hot cores} &   \multicolumn{1}{c}{}                \\ \hline 
S$_2$H$_2$       &  139.885  & 9   & $<$8.5$\times$10$^{11}$ &  $<$4.5$\times$10$^{-11}$  &     &    & \\                       
HSO              &  158.391  & 30  & $<$1.5$\times$10$^{12}$      &  $<$7.9$\times$10$^{-11}$  &     &     &  \\   
HCS              &  161.128  & 30  & $<$3.0$\times$10$^{13}$      &  $<$1.6$\times$10$^{-9}$    &   $\sim$1$-$5$\times$10$^{-10}$ &  &
\citet{Agundez2018}   \\
HSC              &  81.200   & 17   & $<$7.5$\times$10$^{11}$      &  $<$3.9$\times$10$^{-11}$   &   $\sim$6$\times$10$^{-12}$       &  
&  \citet{Agundez2018}  \\
HCCS             &  88.023   & 4   & $<$3.6$\times$10$^{11}$       &  $<$1.9$\times$10$^{-11}$  &     &    &   \\
NCS              &  91.501   & 4   & $<$7.0$\times$10$^{11}$        &   $<$3.7$\times$10$^{-11}$  &     &    &    \\  
HNCS             &  82.102   & 8   & $<$4.2$\times$10$^{11}$      &   $<$2.2$\times$10$^{-11}$   &  $\sim$4$\times$10$^{-12}$    
&    $\sim$8$\times$10$^{-12}$  &  \citet{Adande2010}  \\ 
HSCN             &  92.230   & 5   & $<$4.0$\times$10$^{11}$      &   $<$2.1$\times$10$^{-11}$   &  $\sim$3.5$\times$10$^{-12}$   
&   $\sim$2$\times$10$^{-12}$  &  \citet{Adande2010}   \\ 
HCNS             &  86.118   & 8   & $<$8.0$\times$10$^{10}$      &   $<$4.2$\times$10$^{-12}$       &   &  & \\  
HSNC             &  87.900   & 3   & $<$8.0$\times$10$^{10}$      &   $<$4.2$\times$10$^{-12}$       &   &   & \\
H$_2$C$_3$S      &  85.728   & 8   & $<$8.0$\times$10$^{12}$       &  $<$4.2$\times$10$^{-10}$ &   &  &\\
CH$_3$SH           &  99.185   & 7   & $<$1.2$\times$10$^{13}$       &  $<$6.3$\times$10$^{-10}$  &   &  $\sim$4$\times$10$^{-9}$  &
 \citet{Majumdar2016}   \\ 
CS$^+$           &  103.933  & 7   & $<$1.5$\times$10$^{12}$           &   $<$7.9$\times$10$^{-11}$  &   &  &  \\ 
HSCO$^+$            &  90.183   & 4   & $<$3.5$\times$10$^{11}$      &   $<$1.8$\times$10$^{-11}$  &   &   & \\
HOCS$^+$            &  91.624   & 4   & $<$3.5$\times$10$^{11}$      &   $<$1.8$\times$10$^{-11}$  &   &    & \\ 
HOSO$^+$            &  88.452   & 4   & $<$4.5$\times$10$^{11}$      &   $<$1.8$\times$10$^{-11}$  &   &  &  \\
\hline \hline
\end{tabular}
\end{center}
\noindent
\tablefoot{
(1) The rms has been calculated for a channel width of $\approx$0.33 km s$^{-1}$. 
The obtained rms is similar in the two
surveyed positions. (2) 3$\sigma$ upper limits assuming LTE, 
T$_{rot}$ = 10 K, and a linewidth of 0.6 km s$^{-1}$. Abundances in the Horsehead were derived assuming N(H$\rm _2$) = 1.9$\rm \times ~10^{22}~cm^{-2}$}.
\end{table*}

\begin{table*}[]
\caption{Mean values for kinetic temperatures, gas densities, column densities, and molecular abundances from RADEX models explored with MCMC.}     
\tiny        
\label{Tab:RADEX_CD}      
\centering          
\begin{tabular}{l | cccc | cccc}     
\hline \hline 
              & \multicolumn{4}{c}{PDR} &  \multicolumn{4}{|c}{Core} \\
\hline    
Species & $\rm T_{K} $ & n$\rm _{H_{2}}$ &N(X) & Abundance & $\rm T_{K} $ & n$\rm _{H_{2}}$ &N(X) & Abundance\\
             & (K) & ($\rm cm^{-3}$) & ($\rm cm^{-2}$) & $\rm {N(X)} \over {N(H)+2N(H_{2})} $ & (K) & ($\rm cm^{-3}$) & ($\rm cm^{-2}$) & $\rm {N(X)} \over {N(H)+2N(H_{2})} $ \\
\hline
CS    & 58$\rm \pm$10 & (5.3$\rm \pm$1.7)$\rm \times 10^{4}$ & (1.7$\rm \pm$0.4)$\rm \times 10^{13}$ & (4.5$\rm \pm$1.1)$\rm \times 10^{-10}$ & 27$\rm \pm$6 & (9.5$\rm \pm$3.2)$\rm \times 10^{4}$ & (3.9$\rm \pm$0.7)$\rm \times 10^{13}$ & (6.7$\rm \pm$1.2)$\rm \times 10^{-10}$ \\
SO   & 44$\rm \pm$12 & (6.9$\rm \pm$2.2)$\rm \times 10^{4}$ & (1.5$\rm \pm$0.2)$\rm \times 10^{13}$
& (3.9$\rm \pm$0.5)$\rm \times 10^{-10}$ & 15$\rm \pm$3 & (2.1$\rm \pm$0.7)$\rm \times 10^{5}$ & (3.0$\rm \pm$0.4)$\rm \times 10^{13}$
& (5.2$\rm \pm$0.7)$\rm \times 10^{-10}$ \\
SO$\rm_{2}$ & 58$\rm \pm$10 & (7.0$\rm \pm$2.3)$\rm \times 10^{4}$ & (4.4$\rm \pm$0.8)$\rm \times 10^{12}$ & (1.2$\rm \pm$0.2)$\rm \times 10^{-10}$  & 25$\rm \pm$8 & (1.1$\rm \pm$0.4)$\rm \times 10^{5}$ & (4.3$\rm \pm$0.7)$\rm \times 10^{12}$ & (7.4$\rm \pm$1.2)$\rm \times 10^{-11}$ \\
OCS & 61$\rm \pm$10 & (7.5$\rm \pm$3.7)$\rm \times 10^{4}$ & (1.9$\rm \pm$0.3)$\rm \times 10^{12}$
& (5.0$\rm \pm$0.8)$\rm \times 10^{-11}$ & 24$\rm \pm$6 & (1.0$\rm \pm$0.5)$\rm \times 10^{5}$ & (3.0$\rm \pm$0.5)$\rm \times 10^{12}$
& (5.2$\rm \pm$0.9)$\rm \times 10^{-11}$ \\
o-H$\rm _{2}$CS & 59$\rm \pm$10 & (5.0$\rm \pm$1.7)$\rm \times 10^{4}$ & (9.0$\rm \pm$1.0)$\rm \times 10^{11}$ & (2.4$\rm \pm$0.3)$\rm \times 10^{-11}$ & 18$\rm \pm$6 & (9.1$\rm \pm$3.4)$\rm \times 10^{4}$ & (2.4$\rm \pm$0.3)$\rm \times 10^{12}$ & (4.1$\rm \pm$0.5)$\rm \times 10^{-11}$ \\
p-H$\rm _{2}$CS & -- & -- & -- & -- & 22$\rm \pm$6 & (9.0$\rm \pm$3.4)$\rm \times 10^{4}$ & (9.0$\rm \pm$2.0)$\rm \times 10^{11}$ & (1.6$\rm \pm$0.4)$\rm \times 10^{-11}$ \\
HCS$\rm ^{+}$ & 54$\rm \pm$10 & (3.0$\rm \pm$0.8)$\rm \times 10^{4}$ & (4.1$\rm \pm$0.8)$\rm \times 10^{11}$ & (1.1$\rm \pm$0.2)$\rm \times 10^{-11}$ & -- & -- & -- & --\\
\hline    
\end{tabular}
\tablefoot{Abundances are computed assuming $\rm N(H_{2}) = 1.9 \times 10^{22}~cm^{-2}$ in the PDR and $\rm N(H_{2}) = 2.9 \times 10^{22}~cm^{-2}$ \citep{Gerin2009}.}    
\end{table*}

\section{Rotational diagrams}\label{Sec:rot_diag}
We have used the rotational diagram technique to derive rotational temperatures, $\rm T_{rot}$, and beam-averaged
column densities, N(X). The results are summarized in Table~\ref{Tab:rotational_diagrams}. This  method assumes that the population of the molecular energy levels can be characterized by a single temperature, T$_{\rm rot}$, and that this temperature is uniform along the line of sight. Given the large beam sizes in our observations ($\sim$12$\arcsec$ at 200 GHz to $\sim$25$\arcsec$ at 100 GHz), the assumption of a single T$_{\rm rot}$ for each molecule and position is an over-simplification, yet a valid starting point for our analysis. We also assume that the emission from the different species is filling the beam. While this is most likely the case for many species (see Fig. \ref{Fig:Map}), it might bias our results in some cases.

In its simplest formulation, assuming optically thin emission and the Rayleigh-Jeans limit, the rotation temperature and molecular column density can be derived from the equations

\begin{equation}
ln\left({N_{up}}\over{g_{up}}\right) = ln\left({N_{tot}}\over{Q_{rot}}\right) - \frac{E_{up}}{kT_{rot}}
\end{equation}
\begin{equation}
{\rm N}_{up} (cm^{-2})= 1.94 \times 10^3 \nu^2 (GHz) {\rm W} / {\rm A_{ul}} (s^{-1}) 
,\end{equation}
where $W$ is the velocity-integrated line area in units of ${\rm K~km s}^{-1}$. We assume that the emission is extended compared with the beam size, that is, there is beam filling factor of 1, and  consider only S/N (Area)$\rm >$5 detections to 
compute rotational diagrams.

The assumption of optically thin emission cannot be applied to the abundant species  CS and SO. In these cases, we use the expression

\begin{equation}\label{Eq:rot_diag_opacity}
ln\left({N_{up}^{obs}}\over{g_{up}}\right) = ln\left({N_{tot}^{obs}}\over{Q_{rot}}\right) - \frac{E_{up}}{kT_{rot}} - ln(C_{\tau})
,\end{equation}
with 
\begin{equation}
C_{\tau}=\frac{\tau}{1-e^{-\tau}}
\end{equation}
being the optical depth correction factor \citep{Goldsmith1999}. Line opacities, $\tau$, are derived for each individual line assuming $^{12}$C/$^{13}$C=60 and $^{32}$S/$^{34}$S=22.5 \citep{Wilson1994}.  For a few species, we detected only one transition and we need to assume a certain T$_{\rm rot}$ to estimate column densities. Based on the results of the rotational diagrams of molecules with similar dipole moments   \citep[Table \ref{Tab:rotational_diagrams} and ][]{Fuente2017, Goicoechea2006}, we assume T$_{\rm rot}$=10 K in these cases. The same rotational temperature is used to calculate the 3$\sigma$ column density upper limits shown in Table~\ref{Tab:CD_UL}. 

\section{RADEX models}\label{Sec:RADEX_models}
We used the molecular excitation code RADEX \citep{vanDerTak2007} to get a more detailed description of the physical conditions and molecular abundances at the two positions observed. RADEX is a non-Local Thermodynamic Equilibrium (LTE) molecular excitation and local radiative transfer code that provides line intensities of molecular species for a given combination of $T_{k}$, $n_{H_2}$ , and $N(X)$. We used a Monte Carlo Markov chain (MCMC) methodology to explore the parameter space using a Bayesian inference approach. Specifically, we used the \textit{emcee} \citep{Foreman2012} implementation of the Affine Invariant MCMC Ensemble sampler methods by \cite{Goodman2010}.

When several isotopologues of the same species were detected, as is the case for CS and SO, all the isotopologues were modelled simultaneously assuming the isotopic ratios $\rm ^{13}C/^{12}C=60$ and $\rm ^{34}S/^{32}S=22.5$. Based on our previous knowledge of the region and previous studies by the WHISPER team \citep{Gratier2013}, we decided to assume a Gaussian prior for the kinetic temperature, with mean value and standard deviation $\rm T_{k}=60\pm10$ K in the PDR and $\rm T_{k}=25\pm10$ K in the core, and a log-normal prior for the gas density with mean value and standard deviation $\rm log_{10}(n_{H_{2}} [cm^{-3}])=4.8\pm 0.2$ in the PDR and $\rm log_{10}(n_{H_{2}} [cm^{-3}])=5.0\pm 0.2$ in the core. We assumed a flat prior for column densities between 10$\rm ^{9}~cm^{-2}$ and  10$\rm ^{15}~cm^{-2}$. An overview of the collisional partners and collisional excitation rates used for the different species modelled is shown in Table \ref{Tab:coll_partners_and_rates}. For SO transitions, collisional rates are available either in the range 10 to 50 K \citep{Lique2006B} or 60 to 300 K \citep{Lique2006A}. For our simulations, we used the low temperature collisional rates from \cite{Lique2006B}.

Histograms and density plots showing the results of the MCMC analysis are depicted in Fig. \ref{Fig:MCMC_LVG}. The kinetic temperatures and column densities thus derived are  summarized in Table \ref{Tab:RADEX_CD}. Kinetic temperatures derived from RADEX are systematically higher than those derived from rotational diagrams, which is consistent with sub-thermal excitation. For most molecules, column densities derived from rotational diagrams and RADEX give compatible results, within uncertainties. However, some prominent exceptions are observed. In the PDR, beam-averaged column densities of CS and SO derived from rotational diagrams are two times larger than values derived using RADEX. In the core, the SO abundance derived from rotational diagrams is two times larger than the abundance obtained from RADEX. Since RADEX models take into account non-LTE effects, we consider results from RADEX models to be better constrained than those coming from rotational diagrams, and we adopt them in the following discussion. We note that, because of the effects of high optical depth, column densities derived using LTE should be lower than those obtained using RADEX. However, we have corrected LTE values for opacity effects (see Eq. \ref{Eq:rot_diag_opacity}). The opacity has been derived from the ratio of the integrated intensity of the main isotopologue over the less abundant one, assuming that $T_{ex}$ is the same for all the isotopologues. An overestimation of the opacity would lead to an error in the column density calculation. In addition, the assumption of the same rotation temperature for all transitions used in LTE is not always accurate. In any case, the differences between LTE and RADEX models are within the uncertainties inherent to this kind of calculation.

\begin{figure}[]
\begin{center}
 \includegraphics[scale=0.5,trim = 0mm 0mm 0mm 0mm,clip]{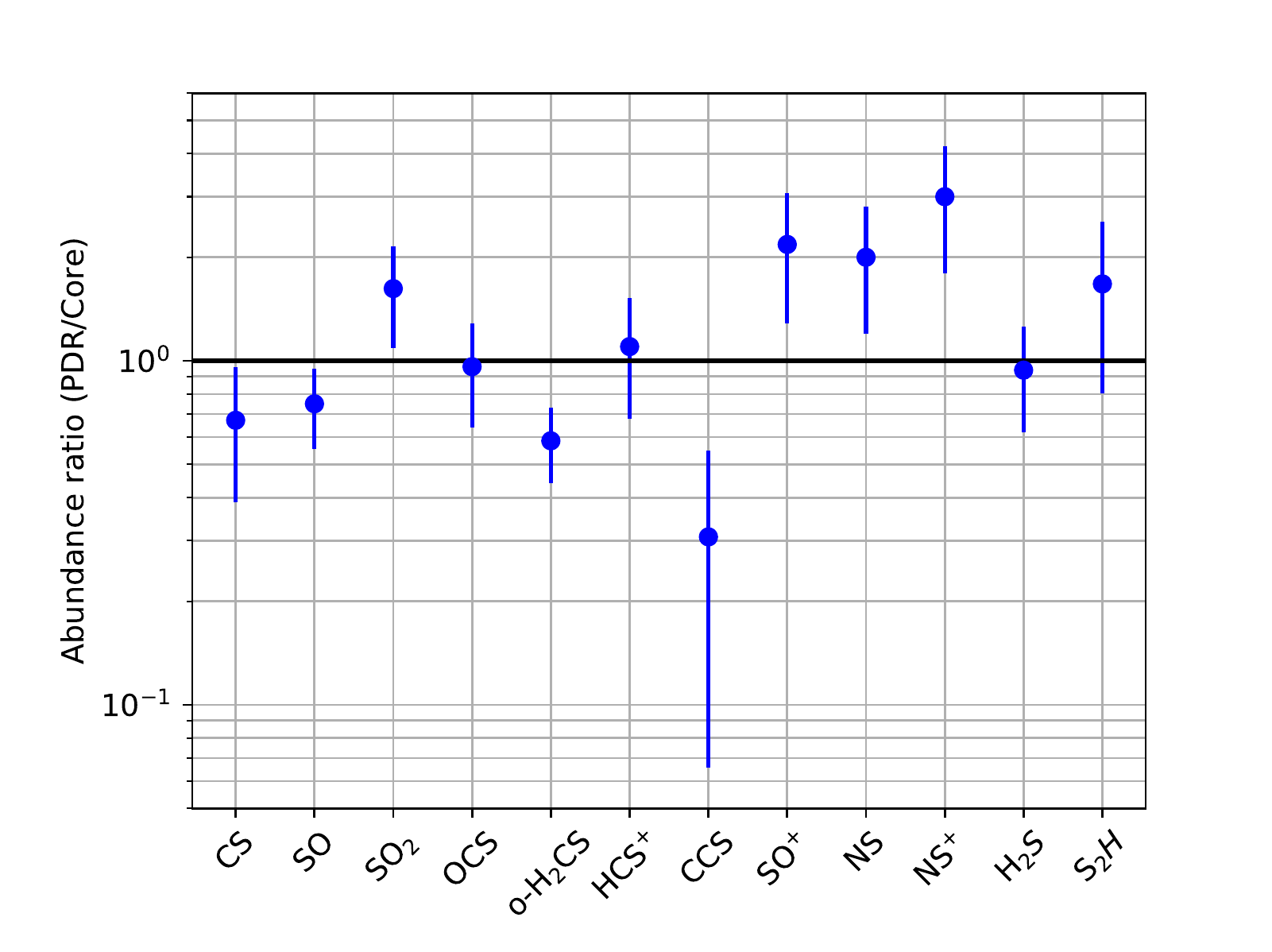} 
 \caption{Ratio between the abundance in the PDR/core position for each species.}
 \label{Fig:Abundance_ratio}
\end{center}
\end{figure}

\section{Molecular abundances}
To derive the abundances shown in Tables \ref{Tab:rotational_diagrams} and \ref{Tab:RADEX_CD} we have assumed  $\rm N(H_{2}) = 1.9 \times 10^{22}~cm^{-2}$ at PDR and $\rm N(H_{2}) = 2.9 \times 10^{22}~cm^{-2}$ at the core \citep{Gerin2009},  and N(H) = 0 at both positions, meaning that gas is mostly molecular. Given the large beam sizes implied, we consider that this is  a valid assumption for the PDR. The $\rm N(H_{2})$ values were derived using a beam of 12$\arcsec$. Following our assumption that the emission is extended and uniform, they should be the same with a larger beam of 25$\arcsec$. We assume this hypothesis to be coherent with our previous studies \citep{Gerin2009b, Pety2012, Guzman2014}.

Our first result is that the fractional abundances and molecular abundance ratios estimated towards the two surveyed positions agree within a factor of approximately three. This gentle gradient is likely due to the limited angular resolution of our observations, which do not resolve the thin outer layers of the PDR. In fact, using interferometric observations, \cite{Goicoechea2006} derived CS abundances towards the outer PDR layers that are approximately five  times larger than those presented in this study. Moreover, the line of sight towards the core also intersects the PDR layer in the cloud surface, which makes the detection of chemical gradients more difficult. A factor of two to three, however, can be significant given the high S/N ratio in our spectra and the peculiar observational strategy of WHISPER, which avoids calibration and pointing errors. The C- bearing species C$_2$S and o-H$\rm _{2}$CS present fractional abundances a factor of $>$ two higher in the core than towards the PDR (see Fig~\ref{Fig:Abundance_ratio}). In contrast, the O-bearing molecules SO, SO$\rm _{2}$ , and OCS present similar abundances towards both positions. The same is true for the HCS$^+$. Other species, S$_2$H, SO$^+$, NS, and the recently discovered ion NS$^+$, are more abundant towards the PDR than towards the core. This work identifies  NS and NS$^+$ as new PDR-like species.

We have unsuccessfully searched for the following sulphuretted compounds: CS$^+$, HCS, HSC, HCCS, NCS, HNCS, HSCN, HCNS, HSNC, HSOC$^+$, HOCS$^+$, H$_2$C$_3$S, and CH$_3$SH. Some of them, HCS, HSC, HNCS, HSCN, and CH$_3$SH, were previously detected in the interstellar medium. The radicals HCS and HSC were recently detected towards the dark cloud L483, with abundances of  a few 10$^{-10}$ and various 10$^{-12}$ respectively \citep{Agundez2018}. HNCS and HSCN were detected in the envelope of Sgr B2 and towards TMC~1 \citep{Adande2010}.  The organic sulphur compound CH$_3$SH has been detected in hot cores (Sgr B2: \citealp{Linke1979}, G327.3-0.6: \citealp{Gibb2000}; Orion: \citealp{Kolesnikova2014}). \citet{Majumdar2016} determined a CH$_3$SH  abundance of a few 10$^{-9}$ towards the hot corino IRAS16293$-$2422. The upper limits obtained towards the Horesehead are not good enough to tightly constrain the abundance of these rare species.

We have detected the deuterated sulphur compound HDCS towards the PDR and core positions. This compound was 
detected by \citet{Marcelino2005} towards the dark cores Barnard 1b and TMC1, with HDCS/H$_2$CS$\sim$0.3,
fully consistent with the values derived towards the Horsehead. However, we have
not detected DCS$^+$ with an upper limit to the DCS$^+$/HCS$^+$ ratio of DCS$^+$/HCS$^+$$<$0.2. 
\citet{Fuente2016} detected DCS$^+$ for the first time towards the dark core
Barnard 1b with  DCS$^+$/HCS$^+$$\sim$0.2. The non-detection of DCS$^+$ in the core position is roughly consistent with the DCS$^+$/HCS$^+$ ratio measured in Barnard 1b. 

\section{Discussion}
We present the abundances of 13 S-bearing species (CS, SO, SO$\rm _{2}$, OCS, H$\rm _{2}$CS -- both ortho and para -- HDCS,
C$\rm _{2}$S, HCS$\rm ^{+}$, SO$\rm ^{+}$, H$_2$S, S$_2$H, NS, and NS$^+$) towards the PDR and core positions in the Horsehead nebula. In particular, we highlight the first detection of NS$^+$ in a PDR environment, and of SO$\rm ^{+}$ in the Horsehead. In the following we discuss the chemistry of these ions and compare the abundances measured towards the Horsehead with those found in other prototypical regions.

\subsection{NS$^+$ and SO$^+$}
\begin{figure}[!t]
\begin{center}

 \includegraphics[trim = 20mm 30mm 0mm 0mm, clip, scale=0.35]{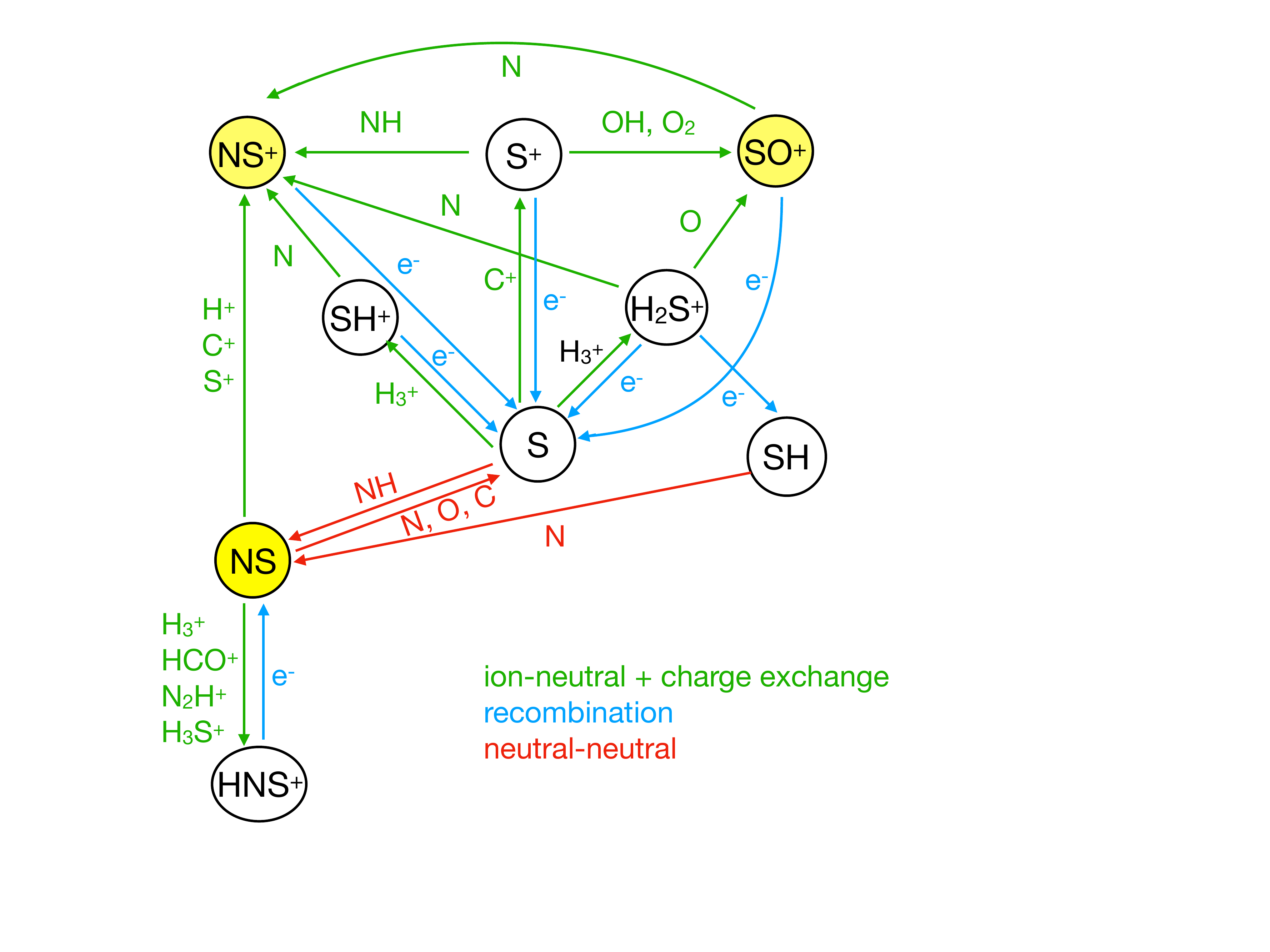}\\
 \caption{Gas phase formation and destruction routes of NS$^+$.}
\label{Fig:NS+}
\end{center}
\end{figure}

The chemistry of NS and NS$^+$ is still poorly understood. \cite{Cernicharo2018} explored the main formation 
mechanisms for these species and their relative importance under  typical physical conditions in dark clouds. We show in Fig.~\ref{Fig:NS+} the formation channels relevant for a PDR environment like that of  the Horsehead nebula, where the S$^+$ + NH  reaction as well as the charge exchange reactions of NS with C$^+$and S$^+$  become formation processes of NS$^+$, as C$^+$ and S$^+$ are the major carbon and  sulfur components. Alternatively, the destruction channel of NS  by atomic carbon stressed in \cite{Cernicharo2018} becomes less critical than in dense cloud conditions.

The NS$^+$ ion is thought to be formed through the reactions

\begin{gather}
{\rm N} + {\rm SH}^+ \rightarrow {\rm NS}^+ + {\rm H} \\
{\rm N} + {\rm SO}^+ \rightarrow {\rm NS}^+ + {\rm O} \\
{\rm S}^+ + {\rm NH} \rightarrow {\rm NS}^+ + {\rm H}  \\ 
{\rm H_{2}S}^+ + {\rm N} \rightarrow {\rm NS}^+ + {\rm H_{2.}}
\end{gather}

These reactions are favoured in UV-illuminated regions. In fact, the ions S$^+$,  SH$^+$, and SO$^+$ are thought to be major gas-phase sulphur reservoirs in the external layers of the dense, high-UV PDRs formed in massive star-forming regions (see e.g. \citealp{Sternberg1995}). The enhanced abundance of SO$^+$ in PDRs has been observationally confirmed by single-dish observations towards prototypical PDRs such as NGC 7023, the Orion Bar, and Mon R2 \citep{Fuente2003,Ginard2012}. The ion SO$^+$ is primarily formed through the reaction
\begin{equation}
\rm
S^+ + OH \rightarrow SO^+ + H 
\end{equation}
and is mainly destroyed by dissociative recombination. 
The radical OH is the product of the endothermic reaction 
\begin{equation}
\rm
O + H_2 \rightarrow OH + H ~( \Delta E/k \sim 3000~K) 
\end{equation}
and  its abundance, and consequently that of SO$^+$, is very sensitive to the gas kinetic temperatures. We have detected for the first time SO$^+$ towards the Horsehead, with N(SO$^+$)/N(SO)$\sim$0.01$-$0.04, where gas kinetic temperatures achieves values of $\sim$100$-$300~K in the most external layers \citep{Habart2005}. This value is only a few times lower than that measured towards the high-UV and denser PDRs Mon R2  and the Orion Bar \citep{Fuente2003, Ginard2012} where the gas kinetic energy might be as high as $\sim$1000~K. The ion SO$^+$ has  been detected with a lower N(SO$^+$)/N(SO) ratio towards the core position, consistent with the presence of a UV-illuminated surface intersecting our line of sight. 

The ion SH$^+$ is only expected to be detectable in warm regions where the endothermic reaction 
\begin{equation}
\rm
S^+ + H_2 \rightarrow  SH^+ + H ~(\Delta E/k \sim 9860~K)
\end{equation}
can proceed \citep{Zanchet2019}. Interestingly, this reaction becomes exothermic  when H$_2$ molecules are in the $v$=2 or higher vibrational levels \citep{Zanchet2013}.  The detection of the SH$^+$ emission lines in the high-mass star-forming region W3~IRS5 \citep{Benz2010} or the Orion Bar \citep{Nagy2013, Muller2014} is interpreted in terms of the reactions of S$^+$ with  vibrationally excited H$_2$ molecules \citep{Zanchet2019}. High angular resolution Atacama Large Millimeter Array (ALMA) images of SH$^+$ and SO$^+$  in the Orion Bar confirm this interpretation \citep{Goicoechea2017}. In these images, the SH$\rm ^{+}$ emission comes only from a thin layer, which is spatially coincident with the H$\rm _{2}$ v=1-0 S(1) emission layer, while SO$\rm ^{+}$ extends deeper into the cloud. The ion SH$^+$ has also been detected in low density environments. In particular, SH$^+$ has been detected in absorption towards Sgr B2 \citep{Menten2011} and in diffuse clouds was observed toward several Galactic sight lines by  \cite{Godard2012}, who showed that turbulent dissipation and shocks provide the required energy to form SH$^+$ in these very low density clouds. To our knowledge, SH$\rm ^{+}$ has not been detected in the Horsehead. This is not surprising since this ion is hardly formed in this nebula. In PDRs, H$\rm _2$($v$=2) and higher vibrational levels are mostly populated by FUV pumping. Since the FUV flux in the Horsehead is not strong, the layer where H$\rm _2$($v$=2) is excited is most likely very narrow, thus strongly diluted to be detected with single-dish telescopes.

Other routes to the formation of  NS$^+$ involve the ionic species  H$_2$S$^+$ and S$^+$. Because of its low ionization potential, sulfur is expected to be ionized up to A$\rm _{v}$ $\rm \sim$ 7 mag (see e.g. \citealp{LeGal2017}) and is a plausible precursor of NS$^+$. The abundance of the ion H$_2$S$^+$, formed by the  reaction S + H$_3^+$, might be important in deeper layers of the molecular cloud depending on the cosmic ray ionization rate \citep{Sternberg1995}.

\begin{figure}[!h]
\begin{center}
 \includegraphics[scale=0.6]{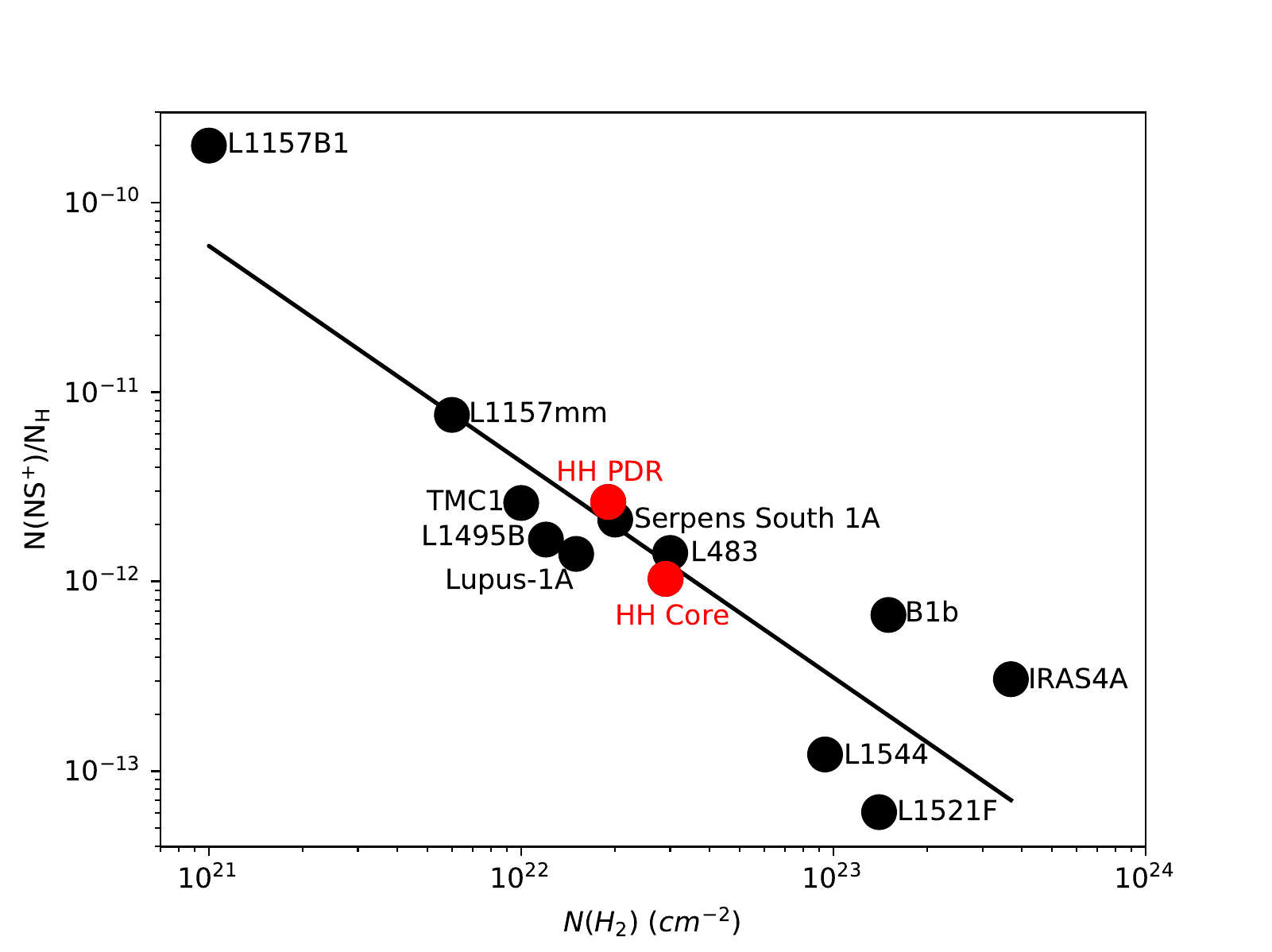}\\
 \caption{Abundances of NS$^+$ as a function of N(H$_2$). Black points corresponds to the detections reported by \citet{Cernicharo2018}. Red points corresponds to the core and PDR positions in the Horsehead. The black solid line depicts a linear fit to the data.}
\label{Fig:corrNS+}
\end{center}
\end{figure}

Within this scenario, NS$^+$ could be considered as a PDR tracer. This ion has been widely detected in the interstellar medium \citep{Cernicharo2018}. One possibility is that NS$^+$ emission is coming from the UV-illuminated external layers of the cloud. To check this hypothesis, in Fig.~\ref{Fig:corrNS+} we represent the NS$^+$ abundance as a function of the total H$_2$ column density for all the reported NS$^+$ detections thus far. The NS$^+$ abundance is anti-correlated with N(H$_2$), as expected if the NS$^+$ emission is coming mainly from the cloud surface. 

The chemistry of NS is poorly understood as well. This compound is detected in very different environments such as dark clouds and hot cores \citep{Fuente2016,Cernicharo2018} as well as towards the Orion Bar \citep{Leurini2006}. Moreover, NS has been detected in comets with an abundance of 0.006-0.12~\% relative to water \citep{Calmonte2016}. Ice evaporation might the origin of NS in hot cores \citep{Vidal2017, Viti2004}. Other desorption mechanisms (photodesorption, cosmic-rays, grain sputtering) might release some NS molecules to the gas phase even in cold environments. We recall that \cite{Fuente2017} also proposed that these desorption processes would release H$\rm_{2}$S and S$\rm_{2}$H molecules from the grain surfaces, boosting their abundances in gas phase.

\begin{table*}[]
\caption{Comparison of fractional abundances with observational templates.}             
\label{Tab:Comps}      
\centering          
\begin{tabular}{l | c | c | c | c }
\hline \hline 
              & \multicolumn{1}{c}{HH-PDR}                 &  \multicolumn{1}{|c}{TMC 1-CP}  &   \multicolumn{1}{|c}{Orion KL}  
              & \multicolumn{1}{|c}{Comets (\% wrt H$_2$O)}   \\
\hline
CS              &  (4.5$\rm \pm$1.1)$\rm \times 10^{-10}$  &  1.4$\rm \times 10^{-9}$  &   7.0$\rm \times 10^{-8}$  &  
0.02$-$0.2  \\

SO              &  (3.9$\rm \pm$0.5)$\rm \times 10^{-10}$  &  7.5$\rm \times 10^{-10}$  &  8.0$\rm \times 10^{-8}$ &
0.04$-$0.3  \\

o-$\rm H_{2}CS$ &  (2.4$\rm \pm$0.3)$\rm \times 10^{-11}$  &  3.0$\rm \times 10^{-10}$ &   5.6$\rm \times 10^{-9}$ &
0.009$-$0.09  \\

OCS             &  (5.0$\rm \pm$0.8)$\rm \times 10^{-11}$  &  1.1$\rm \times 10^{-9}$  &   5.0$\rm \times 10^{-8}$  &
0.03$-$0.4 \\  

CCS             &  (8.0$\rm \pm$2.6)$\rm \times 10^{-12}$  &  3.5$\rm \times 10^{-9}$  &   8.0$\rm \times 10^{-11}$ &  \\ 

SO$\rm _{2}$    &  (1.2$\rm \pm$0.2)$\rm \times 10^{-10}$  &  1.5$\rm \times 10^{-10}$ &   1.9$\rm \times 10^{-7}$ &
0.2  \\

SO$\rm ^{+}$    &  (1.7$\rm \pm$0.5)$\rm \times 10^{-11}$  &                           &                          &    \\

HCS$^+$         &  (1.1$\rm \pm$0.2)$\rm \times 10^{-11}$  &  1.5$\rm \times 10^{-10}$    &                       &  \\

NS                  &  (1.0$\rm \pm$0.2)$\rm \times 10^{-10}$                &  4.0$\rm \times 10^{-10}$    &  3.4$\rm \times 10^{-9}$  & 0.006$-$0.012   \\
NS$^+$          &  (3.0$\rm \pm$0.6)$\rm \times 10^{-12}$                &  2.4$\rm \times 10^{-12}$    &                           &  \\
H$_2$S          &  (3.1$\rm \pm$0.5)$\rm \times 10^{-10}$  &  $<$2.5$\rm \times 10^{-10}$ &  1.5 $\rm \times 10^{-6}$  &
0.13$-$1.5    \\
S$_2$H          &  (8.7$\rm \pm$3.1)$\rm \times 10^{-11}$  &                               &                &         \\ \hline
sulphur budget   &   $\sim$1.5$\times$10$^{-9}$             &   $\sim$8$\times$10$^{-9}$    &    1.9$\times$10$^{-6}$   &  4$\times$10$^{-6}$ (n(H$_2$O)/n$_H$=2.7$\times$10$^{-4}$)\\
\hline \hline   
\end{tabular}
\noindent
\tablefoot{Refs: TMC1: \citet{Agundez2013, Cernicharo2018} and references therein; Orion-KL: \citet{Tercero2010, Esplugues2014, Crockett2014}; Comets: \citet{Calmonte2016, Bockelee2017}.}

\end{table*}

\subsection{Sulphur budget in the Horsehead}
The main reservoir of sulphur in gas phase is still debated. In dark cores, only about 0.05\% is locked in the S-bearing molecules detected at millimeter wavelengths  (see the fractional abundances derived for S-bearing species towards the
dark cloud TMC1-CP in Table~\ref{Tab:Comps}). Unfortunately, most of the plausible major sulphur reservoirs such as S, HS, and S$_2$ are either not observable or very difficult to observe at millimeter wavelengths and one needs to trust to chemical models to estimate the exact amount of sulphur in the gas phase.  It is widely accepted that most of the sulphur is locked in icy grain mantles. This is corroborated by the observations of S-bearing species in comets. In a recent compilation, \citet{Bockelee2017} estimated that the S/O elemental ratio in the comet ice is about 1.5 \%, which is consistent within  a factor of approximately four with the solar value, H$_2$S being the main sulphur reservoir.  Hot cores are compact warm regions where the icy mantles evaporate releasing the sulphur budget to the gas. Orion KL is the best-studied hot core, with full spectral surveys at millimeter and submillimeter wavelengths.  In Table~\ref{Tab:Comps}, we show the abundances of sulphur species that have been derived in this prototypical hot core. Gas phase H$_2$S is the main sulphur reservoir in this hot core and the elemental gas-phase sulphur abundance is $\sim$1.6$\times$10$^{-6}$ \citep{Crockett2014}. \cite{Fuente2019} obtained a similar value for the sulphur gas phase abundance of $\sim$10$^{-6}$ in the edges of the TMC1 molecular cloud.

Based on the WHISPER survey, we have determined a significant inventory of sulphur species and their abundances in the Horsehead PDR.  Only one H$_2$S line has been observed and the estimated value of  the H$_2$S abundance ($\sim$0.3$-$5$\times$10$^{-9}$) is hence uncertain. It seems clear, however, that the abundance of gaseous H$_2$S  is far from that observed in Orion-KL. One possibility is that a large fraction of H$_2$S remains in the solid state. This is expected since the dust  temperature in this PDR is around $\sim$20-30 K in the PDR \citep{Goicoechea2009}, which is below the sublimation temperature of H$_2$S, $\sim$80~K  \citep{JimenezEscobar2011}. The detections of SO$^+$ and NS$^+$ suggest that a significant fraction of sulphur is in the form of S$^+$ in the outer layers of this cloud.  Sulphur chemistry is still poorly known, and especially surface chemistry reactions need further understanding. \cite{Oba2018} demonstrated that chemical desorption is an efficient mechanism for sulphur-bearing-molecule formation at least in some cases, such as H$_2$S. Given the high fraction of sulphur present in a solid state ($\sim$90$\%$), their work highlights the importance of chemical desorption for understanding the abundances of the different molecular species in the gaseous phase.

Although it has been widely debated what the main sulphur reservoirs in the interstellar medium are, little information is available about the sulphur chemistry in PDRs thus far. Our study shows that CS, SO, and H$_2$S are the most abundant observable S-bearing molecules in the Horsehead PDR, with similar abundances within one order of magnitude. The total abundance of S atoms locked in these three species is $\sim$ a few 10$^{-9}$, only $\sim$0.1\% of the total number of atoms. Therefore, they do not constitute the main sulphur reservoir. An important fraction of sulphur atoms might be forming icy mantles in these low-UV irradiated regions. Dust grain temperatures in this low-UV PDR are  $\sim$20-30~K, lower than the evaporation temperature of H$_2$S, the main sulphur reservoir in solid state. In the gas phase, S and S$^+$ are expected to be the main sulphur reservoirs \citep{LeGal2017,Goicoechea2006}. Unfortunately the observation of these species is not easy in the (sub-)millimeter wavelength range. Although weak and difficult to interpret, the detection of a sulphur recombination line by  \citet{Roshi2014}  in NGC~2024 provides evidence for large  abundances of S and S$^+$ in the gas phase.

\section{Conclusions}\label{ref:SumConc}
Photodissociation regions constitute  the transition between diffuse clouds, in which sulphur is mainly in the gaseous phase, and cold cores where S-bearing molecules are heavily depleted; therefore they are essential regions to understand sulphur chemistry. Here we present a complete inventory of S-bearing molecules and their abundances in the Horsehead nebula based on the WHISPER survey. Two positions are targeted in this survey, one coincident with a dense core  and the other at the PDR-line species peak (PDR). The detections towards the PDR are especially interesting since this is the most complete observational study of S-bearing species in a PDR thus far.

We have detected 13 S-bearing species (CS, SO, SO$\rm _{2}$, OCS, H$\rm _{2}$CS -- both ortho and para -- HDCS,
C$\rm _{2}$S, HCS$\rm ^{+}$, SO$\rm ^{+}$, H$_2$S, S$_2$H, NS, and NS$^+$) towards the PDR and core positions in the Horsehead nebula. 

Two prominent results of this study are the detection of SO$^+$ and NS$^+$ for the first time in a PDR. Comparing the core and PDR positions, we find a differentiated chemical behaviour between C-S and O-S bearing species. The C-S bearing species C$_2$S and o-H$\rm _{2}$CS present fractional abundances a factor of $>$ two higher in the core than in the PDR.  In contrast, the O-S bearing molecules  SO, SO$_2$ , and OCS present similar abundances towards both positions. The same is true for the HCS$^+$. A few molecules are more abundant towards the PDR than towards the core: S$_2$H, SO$^+$, NS, and NS$^+$. \cite{Fuente2017} proposed S$\rm _2$H to be a good tracer of low-UV PDRs, where an active grain surface chemistry and gaseous photo-chemistry coexist.  We are aware that, with our limited angular resolution, it is doubtful to conclude that NS$^+$ is mainly coming from the PDR. Yet NS$^+$ behaves like other S-bearing molecules that have been  interpreted as UV tracers, such as SO$^+$ and S$\rm _2$H. Therefore it is worth investigating the link between the enhanced UV-field in PDRs and NS$^ +$ abundances.

\acknowledgements
We thank the Spanish MINECO for funding support from AYA2016-75066-C2-1/2-P, AYA2017-85111-P and ERC under ERC-2013-SyG, G. A. 610256 NANOCOSMOS. JRG thanks the Spanish MCIU for funding support under grant AYA2017-85111-P. This work was supported by the Programme National $``$Physique et Chimie du Milieu Interstellaire$''$ (PCMI) of CNRS/INSU with INC/INP co-funded by CEA and CNES.

\bibliographystyle{aa} 
\bibliography{biblio.bib}

\appendix

\section{Tables and figures}

\begin{table*}[!h]
\caption{Gaussian fits to the IRAM 30m lines.}
\label{Tab:lineFluxes}            
\centering
\resizebox{18cm}{!}{
\begin{tabular}{ll|cccc|cccc}
\hline \hline 
\multicolumn{2}{c|}{}  & \multicolumn{4}{c|}{PDR} &  \multicolumn{4}{c}{Core}  \\ 
\multicolumn{2}{c|}{}  & \multicolumn{1}{c}{T$_{\rm MB}$} & \multicolumn{1}{c}{v$_{\rm LSR}$} & 
\multicolumn{1}{c}{$\Delta$v} & \multicolumn{1}{c|}{$\rm \int T_{MB}$d$v$} & 
\multicolumn{1}{c}{T$_{MB}$}  & \multicolumn{1}{c}{v$_{\rm LSR}$} & \multicolumn{1}{c}{$\Delta$v} & 
\multicolumn{1}{c}{$\int T_{MB}$d$v$} \\
\multicolumn{2}{c|}{}  &  \multicolumn{1}{c}{(K)}  &  \multicolumn{2}{c}{(km $\rm s^{-1}$)}   &
\multicolumn{1}{c|}{(K km $\rm s^{-1}$)} & \multicolumn{1}{c}{(K)}  &  \multicolumn{2}{c}{(km $\rm s^{-1}$)}   &
\multicolumn{1}{c}{(K km $\rm s^{-1}$)} \\
\hline

CS & 2$\rightarrow$1 & 3.00 & 10.70$\pm$0.01 & 0.76$\pm$0.01 & 2.440$\pm$0.003 & 
                                       5.60 & 10.67$\pm$0.01 & 0.79$\pm$0.01 & 4.752$\pm$0.004 \\
      & 3$\rightarrow$2 & 2.20 & 10.67$\pm$0.01 &  0.74$\pm$0.01 & 1.705$\pm$0.009 & 
                                       4.70 & 10.63$\pm$0.01 &  0.74$\pm$0.01 & 3.662$\pm$0.007 \\
      & 5$\rightarrow$4 & 0.94 & 10.65$\pm$0.01 &  0.59$\pm$0.01 & 0.592$\pm$0.007 & 
                                     1.20 &  10.61$\pm$0.01 & 0.63$\pm$0.01 & 0.805$\pm$0.007 \\
\hline
C$^{34}$S & 2$\rightarrow$1 & 0.41 & 10.72$\pm$0.01 & 0.55$\pm$0.01 & 0.241$\pm$0.003 & 
                                                   0.78 & 10.70$\pm$0.01 & 0.61$\pm$0.01 & 0.505$\pm$0.002 \\
                  & 3$\rightarrow$2 & 0.34  & 10.72$\pm$0.01   & 0.54$\pm$0.03 & 0.195$\pm$0.008  & 
                                                    0.49  & 10.67$\pm$0.01   & 0.56$\pm$0.01 & 0.291$\pm$0.007  \\
                  & 5$\rightarrow$4 & 0.05 & 10.57$\pm$0.06   & 0.46$\pm$0.12 & 0.027$\pm$0.006 & - & - & - & - \\
\hline
$^{13}$CS & 2$\rightarrow$1 &  0.11 & 10.71$\pm$0.01  & 0.52$\pm$0.03 & 0.063$\pm$0.003 & 
                                                     0.20 & 10.70$\pm$0.01  & 0.58$\pm$0.02 & 0.127$\pm$0.003 \\
                  & 3$\rightarrow$2 &  0.05 & 10.61$\pm$0.07  &  0.66$\pm$0.16 & 0.037$\pm$0.008 & 
                                                     0.16 & 10.62$\pm$0.01  &  0.51$\pm$0.04 & 0.088$\pm$0.006 \\
\hline
C$^{33}$S & 2$\rightarrow$1 & 0.07  &  11.50$\pm$0.02    & 0.48$\pm$0.05  &   0.034$\pm$0.003 &
                                                   0.09   & 11.46$\pm$0.02    &  0.74$\pm$0.05  &  0.074$\pm$0.004  \\
                  & 3$\rightarrow$2  &  - & - & - & - & 0.08 & 10.92$\pm$0.05   &  0.56$\pm$0.09   &  0.046$\pm$0.007 \\
\hline
CCS & $ 6_{7}$$\rightarrow$$ 5_{6}$ & 0.09 & 10.76$\pm$0.04 & 0.42$\pm$0.09 & 0.042$\pm$0.008 & 
                                                                0.31 & 10.70$\pm$0.01 & 0.44$\pm$0.03 & 0.146$\pm$0.007 \\        
        & $ 7_{6}$$\rightarrow$$ 6_{5}$ & 0.04 & 10.76$\pm$0.03 &  0.35$\pm$0.06 & 0.014$\pm$0.002 & 
                                                                0.09 & 10.70$\pm$0.01 &  0.37$\pm$0.04 & 0.037$\pm$0.003 \\
        & $ 7_{7}$$\rightarrow$$ 6_{6}$ & 0.04 & 10.73$\pm$0.03 &  0.40$\pm$0.07 & 0.017$\pm$0.003 & 
                                                                0.08 & 10.68$\pm$0.01 &  0.47$\pm$0.03 & 0.042$\pm$0.002 \\
        & $ 7_{8}$$\rightarrow$$ 6_{7}$ & 0.07 & 10.71$\pm$0.02 & 0.48$\pm$0.03 & 0.036$\pm$0.002 & 
                                                                0.23 & 10.69$\pm$0.01 & 0.47$\pm$0.01 & 0.117$\pm$0.002 \\
        & $ 8_{7}$$\rightarrow$$ 7_{6}$ & - & - & - & -  & 0.08 & 10.76$\pm$0.02 & 0.33$\pm$0.05 & 0.030$\pm$0.003 \\
        & $ 8_{8}$$\rightarrow$$ 7_{7}$ & 0.04 & 10.74$\pm$0.03 & 0.32$\pm$0.08 & 0.013$\pm$0.002 & 
                                                                0.07 & 10.68$\pm$0.02 & 0.39$\pm$0.05 & 0.031$\pm$0.003 \\                        
\hline
OCS & 7$\rightarrow$6 & 0.02 & 10.72$\pm$0.06  & 0.57$\pm$0.11 & 0.014$\pm$0.003 &
                                          0.06 & 10.69$\pm$0.04  & 0.75$\pm$0.09 & 0.045$\pm$0.005 \\
         & 8$\rightarrow$7 & 0.04 & 10.76$\pm$0.02 & 0.40$\pm$0.04 & 0.019$\pm$0.002  &
                                          0.05 & 10.62$\pm$0.02 & 0.61$\pm$0.05 & 0.033$\pm$0.002  \\
         & 9$\rightarrow$8 & 0.05 & 10.66$\pm$0.05  & 0.48$\pm$0.14 & 0.024$\pm$0.005  &
                                          0.07 & 10.76$\pm$0.03  & 0.45$\pm$0.01 & 0.034$\pm$0.004  \\
         & 11$\rightarrow$10 & - & - & - & - & 
                                          0.05 & 10.60$\pm$0.04 &  0.41$\pm$0.09 & 0.023$\pm$0004\\ 
         & 12$\rightarrow$11 & - & - & - & - &
                                          0.08 & 10.50$\pm$0.05 &  0.59$\pm$0.14 & 0.048$\pm$0.009 \\ 

\hline
p-H$_{2}$CS & $3_{03}$$\rightarrow$$2_{02}$ & 0.08 & 10.72$\pm$0.02 & 0.52$\pm$0.04 & 0.046$\pm$0.003& 
                                                                                 0.23 & 10.65$\pm$0.01 & 0.55$\pm$0.02 & 0.137$\pm$0.003 \\
                                                                                            
                      & $ 4_{04}$$\rightarrow$$3_{03}$ & 0.05 & 10.78$\pm$0.03  & 0.30$\pm$0.08 & 0.017$\pm$0.004 & 
                                                                                       0.23 & 10.68$\pm$0.02  & 0.46$\pm$0.03 & 0.113$\pm$0.005 \\      
                                                                                                                                                                             
                      & $ 6_{06}$$\rightarrow$$5_{05}$ & - & - & - & - &
                                                                                0.04 &  10.76$\pm$0.05  & 0.70$\pm$0.11 & 0.032$\pm$0.005 \\
\hline
o-H$_{2}$CS & $ 3_{13}$$\rightarrow$$2_{12}$ & 0.13 & 10.68$\pm$0.01 & 0.47$\pm$0.03 & 0.066$\pm$0.003 & 
                                                                                  0.33 & 10.65$\pm$0.01 & 0.51$\pm$0.01 & 0.179$\pm$0.003 \\
                                                                                            
                       & $ 3_{12}$$\rightarrow$$2_{11}$ & 0.14 & 10.75$\pm$0.05  &  0.47$\pm$0.03 & 0.071$\pm$0.003 & 
                                                                                  0.30 & 10.71$\pm$0.01  & 0.55$\pm$0.01 & 0.177$\pm$0.003 \\
                                                                                           
                       & $ 4_{14}$$\rightarrow$$3_{13}$ & 0.16 & 10.65$\pm$0.02 & 0.49$\pm$0.04 & 0.084$\pm$0.005 & 
                                                                                  0.31 & 10.61$\pm$0.01 & 0.52$\pm$0.02 & 0.173$\pm$0.005 \\
                                                                                           
                       & $ 4_{13}$$\rightarrow$$3_{12}$ & 0.10 & 10.84$\pm$0.02 & 0.44$\pm$0.05 & 0.046$\pm$0.005 & 
                                                                                  0.28 &  10.80$\pm$0.02 & 0.43$\pm$0.02 & 0.131$\pm$0.005 \\
\hline
HDCS & $3_{03}$$\rightarrow$$2_{02}$ & - & - & - & - &
                                                                        0.06 & 10.63$\pm$0.02 & 0.34$\pm$0.04 & 0.022$\pm$0.002 \\ 
           & $3_{13}$$\rightarrow$$2_{12}$ & - & - & - & - &
                                                                        - & - & - & - \\
            & $3_{12}$$\rightarrow$$2_{11}$ &  0.05 & 10.79$\pm$0.02 & 0.31$\pm$0.05 & 0.018$\pm$0.002 &
                                                                        0.05 & 10.79$\pm$0.02 & 0.50$\pm$0.08 & 0.027$\pm$0.003 \\                 
\hline
HCS$^{+}$ & 2$\rightarrow$1 & 0.13 & 10.75$\pm$0.02 & 0.73$\pm$0.03 & 0.102$\pm$0.004 & 
                                                          0.11 & 10.75$\pm$0.01 & 0.66$\pm$0.03 & 0.073$\pm$0.003 \\
                         & 5$\rightarrow$4 & 0.11 & 10.75$\pm$0.03  & 0.62$\pm$0.07 & 0.069$\pm$0.007 &
                                                             - & - & - & - \\
                         & 6$\rightarrow$5 & 0.06 &  10.64$\pm$0.05  & 0.65$\pm$0.14 & 0.039$\pm$0.006 &
                                                             - & - & - & - \\
\hline
SO & $ 2_{2}$$\rightarrow$$ 1_{1}$ & 0.45 & 10.72$\pm$0.01 & 0.50$\pm$0.01 & 0.241$\pm$0.005 & 
                                                              0.50 & 10.66$\pm$0.01 & 0.55$\pm$0.08 & 0.293$\pm$0.003  \\
      & $ 2_{3}$$\rightarrow$$ 1_{2}$ & 3.20 & 10.73$\pm$0.01 & 0.58$\pm$0.01 & 2.004$\pm$0.003 &
                                                              4.50 & 10.66$\pm$0.01 & 0.65$\pm$0.01 & 3.151$\pm$0.003 \\
      & $ 3_{2}$$\rightarrow$$ 2_{1}$ & 0.55 & 10.73$\pm$0.01 &  0.48$\pm$0.01 & 0.285$\pm$0.004 & 
                                                              0.73 & 10.64$\pm$0.01 &  0.50$\pm$0.01 & 0.392$\pm$0.005 \\
      & $ 3_{3}$$\rightarrow$$ 2_{2}$ & 0.70 & 10.72$\pm$0.01 & 0.43$\pm$0.01 & 0.317$\pm$0.005 & 
                                                              1.10 & 10.61$\pm$0.01 & 0.45$\pm$0.01 & 0.521$\pm$0.005 \\
      & $ 3_{4}$$\rightarrow$$ 2_{3}$ & 2.40 & 10.68$\pm$0.01 & 0.56$\pm$0.01 & 1.466$\pm$0.005 & 
                                                              4.40 & 10.61$\pm$0.01 & 0.58$\pm$0.01 & 2.731$\pm$0.007 \\
      & $ 4_{3}$$\rightarrow$$ 3_{2}$ & 0.55 & 10.69$\pm$0.01 & 0.52$\pm$0.02 & 0.305$\pm$0.011 & 
                                                              0.92 & 10.60$\pm$0.01 & 0.45$\pm$0.02 & 0.442$\pm$0.013 \\
      & $ 5_{4}$$\rightarrow$$ 4_{3}$ & 0.34 & 10.69$\pm$0.01 & 0.63$\pm$0.01 & 0.225$ \pm $0.003 & 
                                                              0.40 & 10.61$\pm$0.01 & 0.59$\pm$0.02 & 0.254$\pm$0.007 \\
      & $ 5_{5}$$\rightarrow$$ 4_{4}$ & 0.34 & 10.70$\pm$0.01 & 0.45$\pm$0.01 & 0.166$\pm$0.007 & 
                                                              0.40 & 10.59$\pm$0.01 & 0.48$\pm$0.02 & 0.204$\pm$0.007 \\
      & $ 5_{6}$$\rightarrow$$ 4_{5}$ & 1.23 & 10.68$\pm$0.01 & 0.62$\pm$0.01 & 0.812$\pm$0.007 & 
                                                              1.48 & 10.62$\pm$0.01 & 0.60$\pm$0.01 & 0.944$\pm$0.006 \\
\hline

$ ^{34}$SO & $ 2_{2}$$\rightarrow$$ 1_{1}$ & - & - & - & - & 
                                                                       0.02 & 10.82$\pm$0.06 & 0.99$\pm$0.12 & 0.022$\pm$0.002 \\
                         & $ 2_{3}$$\rightarrow$$ 1_{2}$ & 0.22 & 10.74$\pm$0.01 &  0.54$\pm$0.02 & 0.130$\pm$0.003 & 
                                                                       0.30 & 10.65$\pm$0.01  & 0.54$\pm$0.02 & 0.172$\pm$0.003 \\
                         & $ 3_{2}$$\rightarrow$$ 2_{1}$ &  - & - & -  & - & 
                                                                       0.06 & 10.56$\pm$0.03 & 0.38$\pm$0.07 & 0.024$\pm$0.004 \\
                         & $ 3_{4}$$\rightarrow$$ 2_{3}$ & 0.22 & 10.71$\pm$0.01 & 0.46$\pm$0.03 & 0.106$\pm$0.005 & 
                                                                                 0.31 & 10.59$\pm$0.03 & 0.49$\pm$0.02 & 0.164$\pm$0.006 \\
                         & $ 5_{6}$$\rightarrow$$ 4_{5}$ & 0.10 & 10.64$\pm$0.03 & 0.40$\pm$0.11 & 0.043$\pm$0.007 & 
                                                                                 0.07 & 10.50$\pm$0.06 & 0.97$\pm$0.14 & 0.069$\pm$0.008 \\
S$ ^{18}$O & $ 2_{3}$$\rightarrow$$ 1_{2}$ &  - & - & - & - &
                                                                        0.04 & 10.67$\pm$0.04 & 0.66$\pm$0.09 & 0.028$\pm$0.003 \\
\hline 

SO$ ^{+}$ & {\scriptsize 9/2$\rightarrow$7/2 (e)} & 0.10 & 10.70$\pm$0.04 & 0.77$\pm$0.10 & 0.082$\pm$0.009 & 
                                                                                 0.07 & 10.66$\pm$0.05 & 0.70$\pm$0.10 & 0.050$\pm$0.006 \\
                 & {\scriptsize 9/2$\rightarrow$7/2 (f)} & 0.09 & 10.73$\pm$0.03 & 0.67$\pm$0.07 & 0.068$ \pm 0.006$  & 
                                                                                    0.05 & 10.77$\pm$0.11 & 0.96$\pm$0.30 & 0.047 $ \pm 0.011$ \\
                 & {\scriptsize 11/2$\rightarrow$9/2 (e)} & 0.07  & 10.69$\pm$0.04 & 0.50$\pm$0.08 & 0.037$ \pm 0.006$  & 
                                                                                  0.05  & 10.78$\pm$0.07 & 0.48$\pm$0.16  & 0.023$ \pm 0.006$\\
                 & {\scriptsize 11/2$\rightarrow$9/2 (f)} & 0.05 & 10.72$\pm$0.05 & 0.67$\pm$0.12 & 0.037$ \pm 0.005$  & 
                                                                             - & - & - & - \\
\hline

SO$ _{2}$ & $ 3_{13}$$\rightarrow$$ 2_{02}$ & 0.29 & 10.74$\pm$0.01 & 0.47$\pm$0.02  & 0.143$\pm$0.003 & 
                                                                             0.25 & 10.65$\pm$0.01  & 0.58$\pm$0.01 & 0.156$\pm$0.002 \\
                       & $ 5_{15}$$\rightarrow$$ 4_{04}$ & 0.27 & 10.72$\pm$0.01 & 0.47$\pm$0.03 & 0.132$\pm$0.005 & 
                                                                                   0.28 & 10.62$\pm$0.01  & 0.48$\pm$0.02 & 0.148$\pm$0.004 \\
                       & $ 3_{22}$$\rightarrow$$ 2_{11}$ & 0.09 &  10.70$\pm$0.06 & 0.36$\pm$0.30 & 0.033$\pm$0.007 & 
                                                                                  0.09  &  10.73$\pm$0.04 & 0.68$\pm$0.08 & 0.068$\pm$0.007 \\
                       & $ 4_{22}$$\rightarrow$$ 3_{13}$ & 0.09 &  10.70$\pm$0.03 & 0.58$\pm$0.06 & 0.054$\pm$0.005 & 
                                                                                   0.06 &  10.50$\pm$0.05 & 0.815$\pm$0.153 & 0.049$\pm$0.007 \\
                       & $ 5_{24}$$\rightarrow$$ 4_{13}$ & - & - & - & - & 
                                                                                   0.08   & 10.69$\pm$0.05 & 0.46$\pm$0.11 & 0.039$\pm$0.008 \\
\hline
NS & {\scriptsize 5/2,7/2$\rightarrow$3/2,5/2 (e)} & 0.18 &  10.77$\pm$0.04 &  0.61$\pm$0.08 & 0.12$\pm$0.01 & 
                                                                                  0.20 &  10.91$\pm$0.03 &  0.54$\pm$0.05 & 0.110$\pm$0.010\\
     &  {\scriptsize 5/2,5/2$\rightarrow$3/2,3/2 (e)} & 0.14 &  10.90$\pm$0.04 & 0.51$\pm$0.10 & 0.075$\pm$0.012 & 
                                                                                  0.11 &  10.83$\pm$0.06 & 0.53$\pm$0.21 & 0.059$\pm$0.015 \\
     &  {\scriptsize 5/2,3/2$\rightarrow$3/2,1/2 (e)} & 0.12 &  11.07$\pm$0.04  & 0.59$\pm$0.09 & 0.074$\pm$0.010 & 
                                                                                 0.10 &  10.84$\pm$0.06  & 0.58$\pm$0.22 & 0.063$\pm$0.015  \\
     & {\scriptsize 5/2,7/2$\rightarrow$3/2,5/2 (f)} & 0.24 &  10.50$\pm$0.02 &  0.37$\pm$0.05 & 0.094$\pm$0.012 & 
                                                                               0.20 & 10.34$\pm$0.04  &  0.59$\pm$0.11 & 0.127$\pm$0.018\\
     &  {\scriptsize 5/2,5/2$\rightarrow$3/2,3/2 (f)} & 0.14 &  10.58$\pm$0.07 &  0.56$\pm$0.22 & 0.084$\pm$0.024 & 
                                                                                 0.16 &  10.38$\pm$0.04 &   0.27$\pm$0.12 & 0.045$\pm$0.014 \\
     &  {\scriptsize 5/2,3/2$\rightarrow$3/2,1/2 (f)} & 0.21 &  10.52$\pm$0.03  & 0.31$\pm$0.06  & 0.069$\pm$0.012 & 
                                                                                0.14 &  10.39$\pm$0.06 & 0.45$\pm$0.16  & 0.017$\pm$0.019  \\
\hline                                                                                                                      
NS$^+$ & 2$\rightarrow$1 & 0.034 & 10.46$\pm$0.04 &  0.77$\pm$0.08 & 0.028$\pm$0.002 & 
                                              0.017 & 10.44$\pm$0.09 &  1.0$\pm$0.3 & 0.019$\pm$0.004\\
              & 3$\rightarrow$2 & 0.035 & 10.64$\pm$0.10 & 0.76$\pm$0.26 & 0.028$\pm$0.007 & 
                                               -- & -- & -- & --\\
                             
\hline \hline
\end{tabular}
}
\tablefoot{Tabulated errors are the numerical errors of the Gaussian fitting. Other kinds of errors such as calibration or pointing errors are not considered.}    
\end{table*}

\begin{figure*}[!h]
\begin{center}
 \includegraphics[scale=0.44,trim = 36mm 40mm 0mm 0mm,clip]{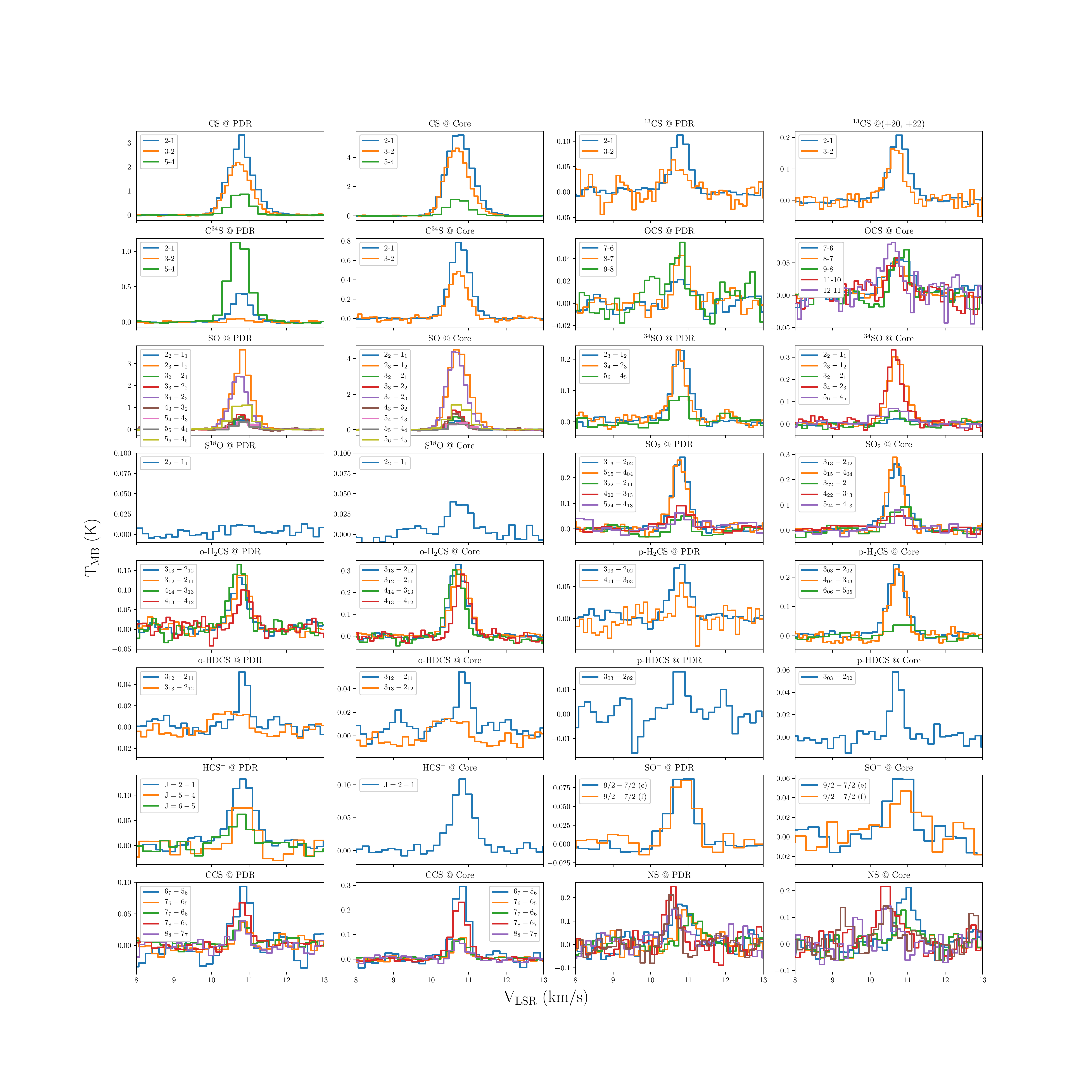}\\
 \caption{Detected species in the Horsehead.}
\label{Fig:allDetections}
\end{center}
\end{figure*}

\section{Distribution of model parameters from MCMC simulations}

\begin{figure*}[!t]
\begin{center}
 \includegraphics[trim=6mm 6mm 6mm 8mm, clip, scale=0.5]{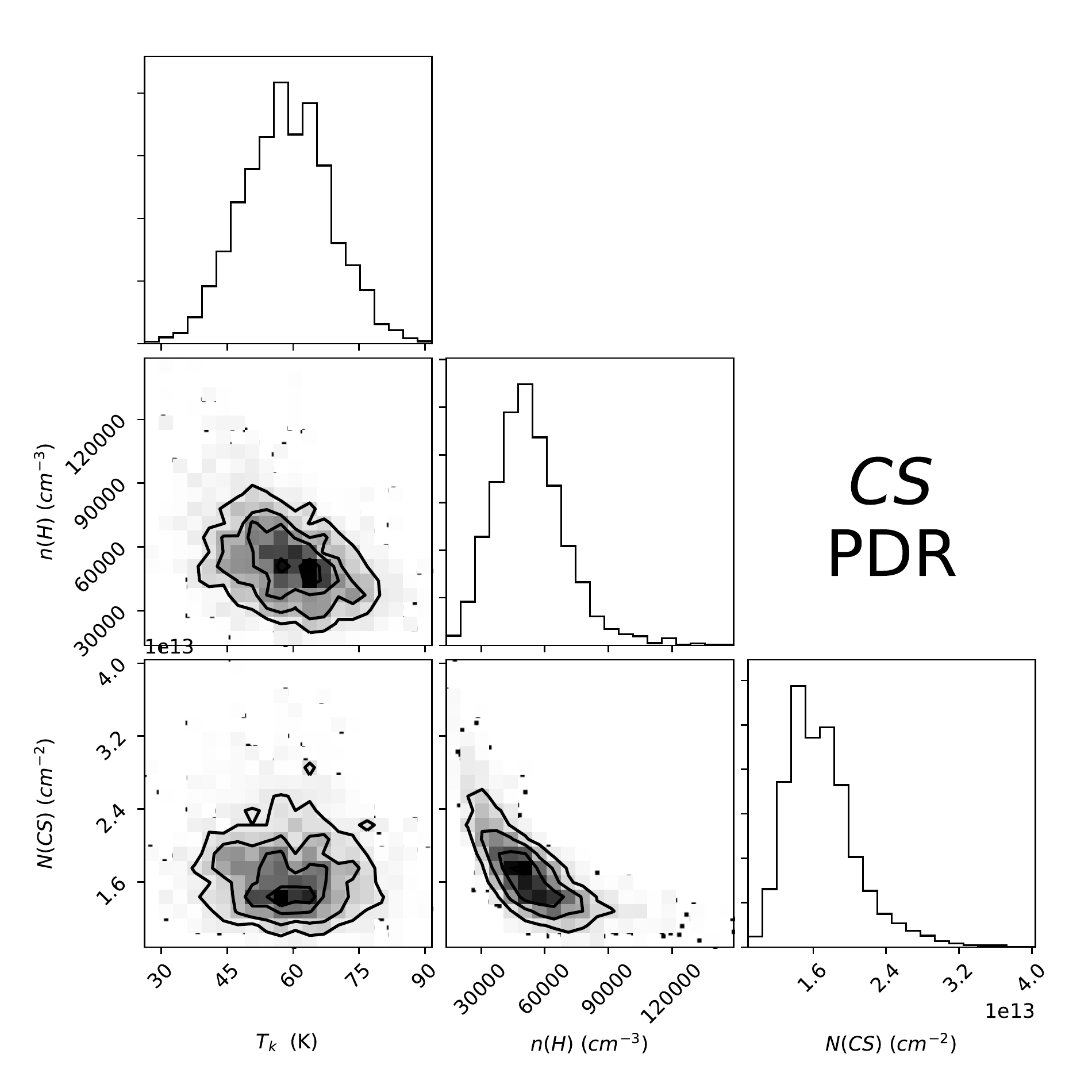}\includegraphics[trim=6mm 6mm 0mm 8mm, clip, scale=0.5]{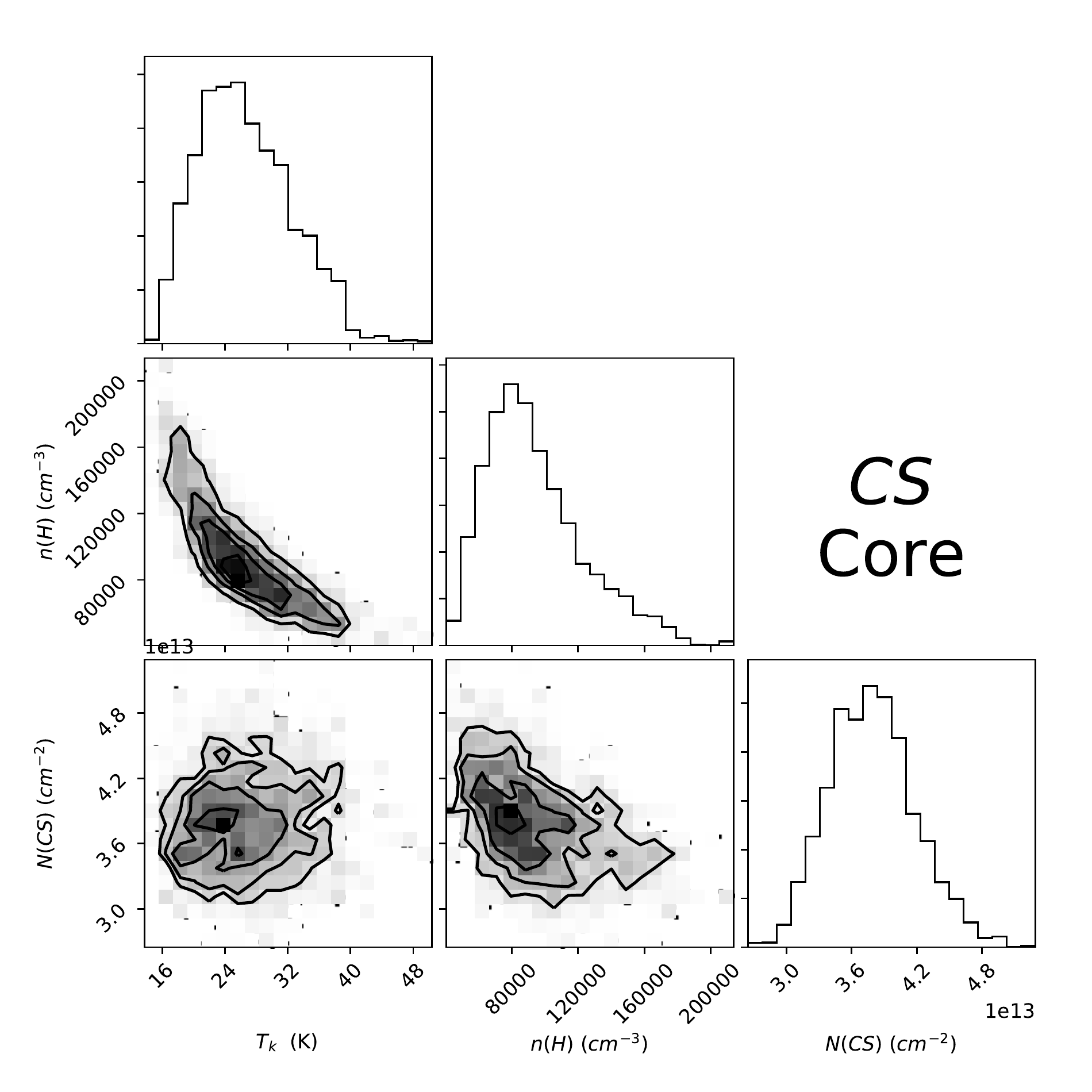}\\
  \includegraphics[trim=6mm 6mm 6mm 8mm, clip, scale=0.5]{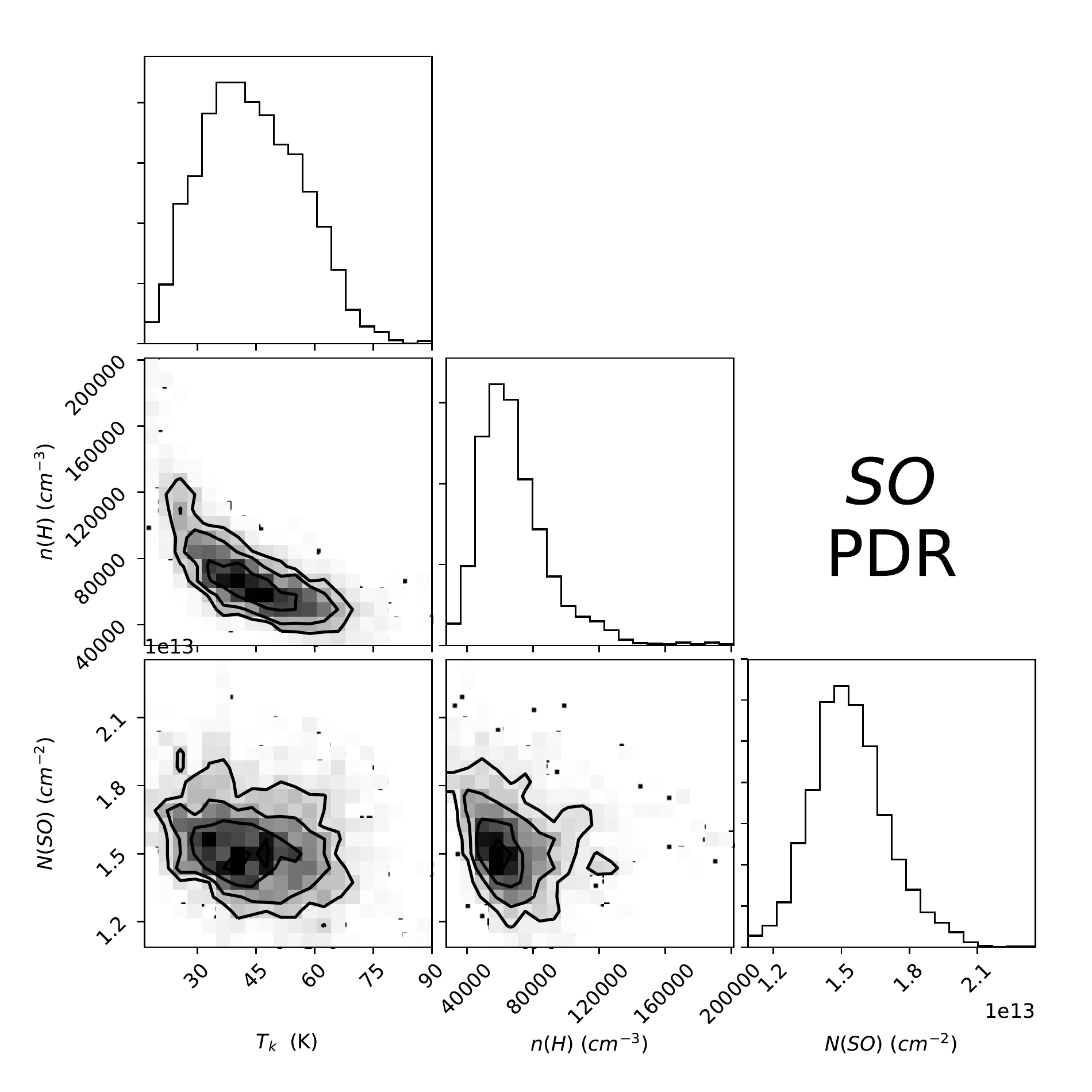}\includegraphics[trim=6mm 6mm 0mm 8mm, clip, scale=0.5]{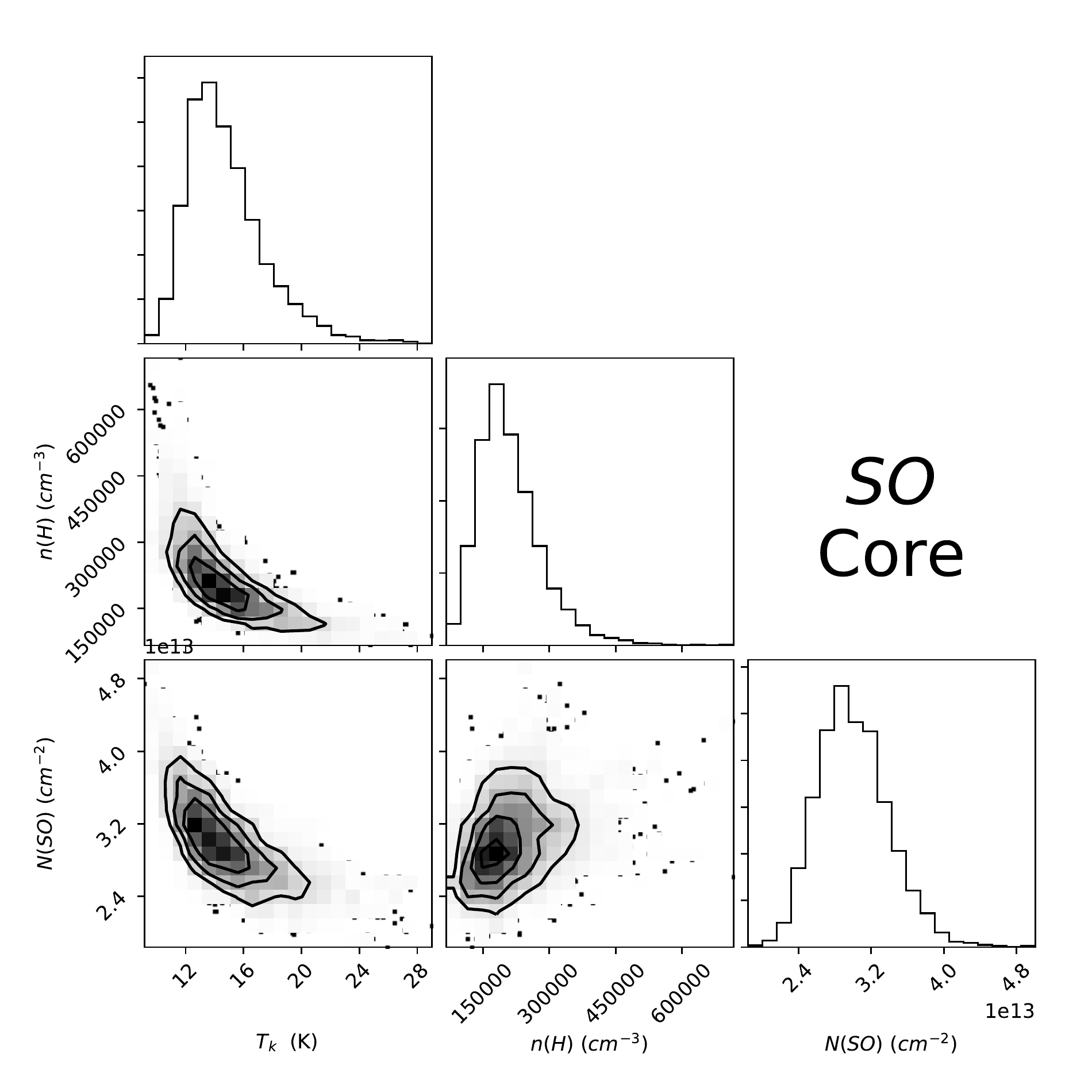}\\
\caption{Histograms of the MCMC RADEX parameter distribution for the different molecules and regions studied.}
\label{Fig:MCMC_LVG}
\end{center}
\end{figure*}
 
\addtocounter{figure}{-1}
\begin{figure*}[!h]
\begin{center}
 \includegraphics[trim=6mm 6mm 6mm 8mm, clip, scale=0.5]{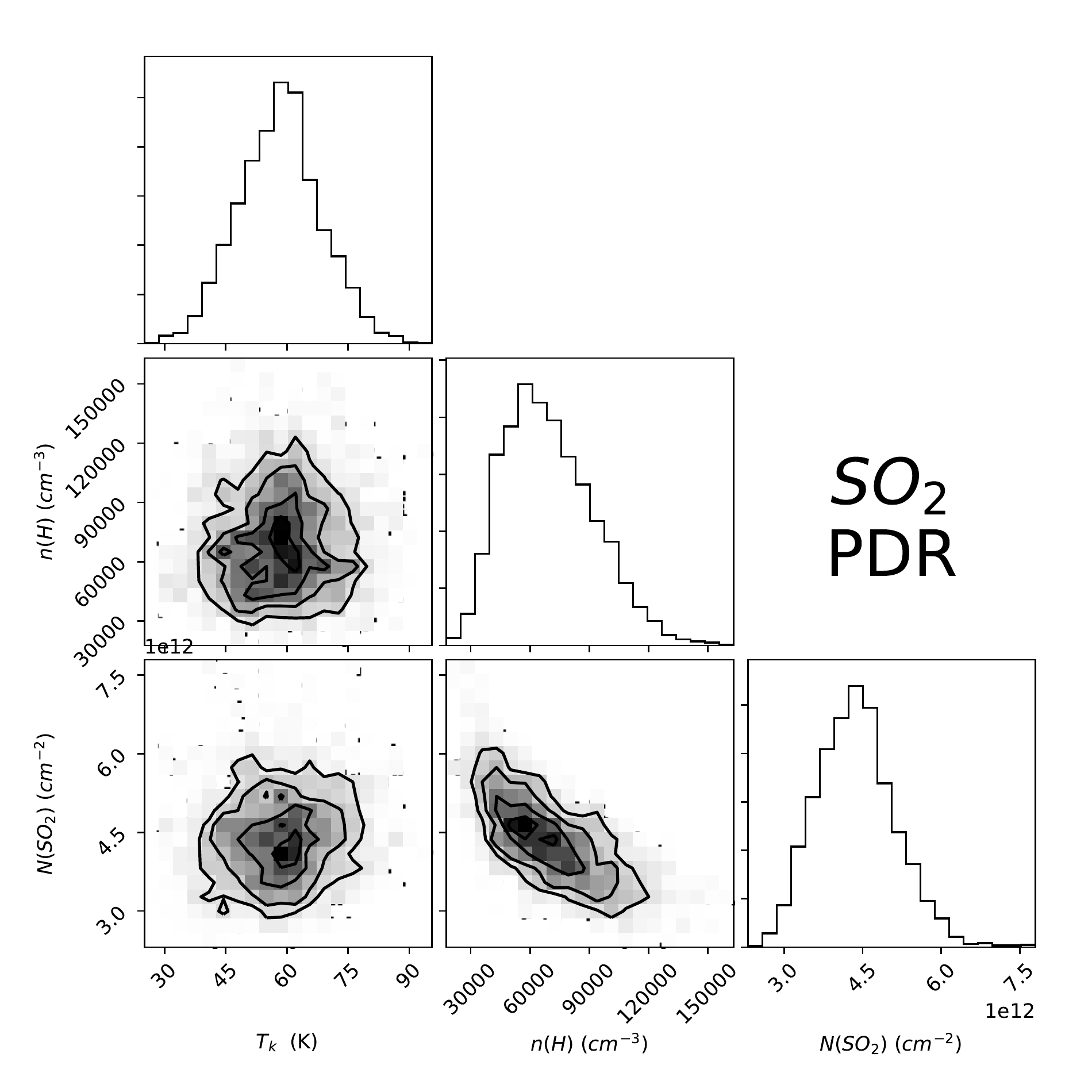}\includegraphics[trim=6mm 6mm 0mm 8mm, clip, scale=0.5]{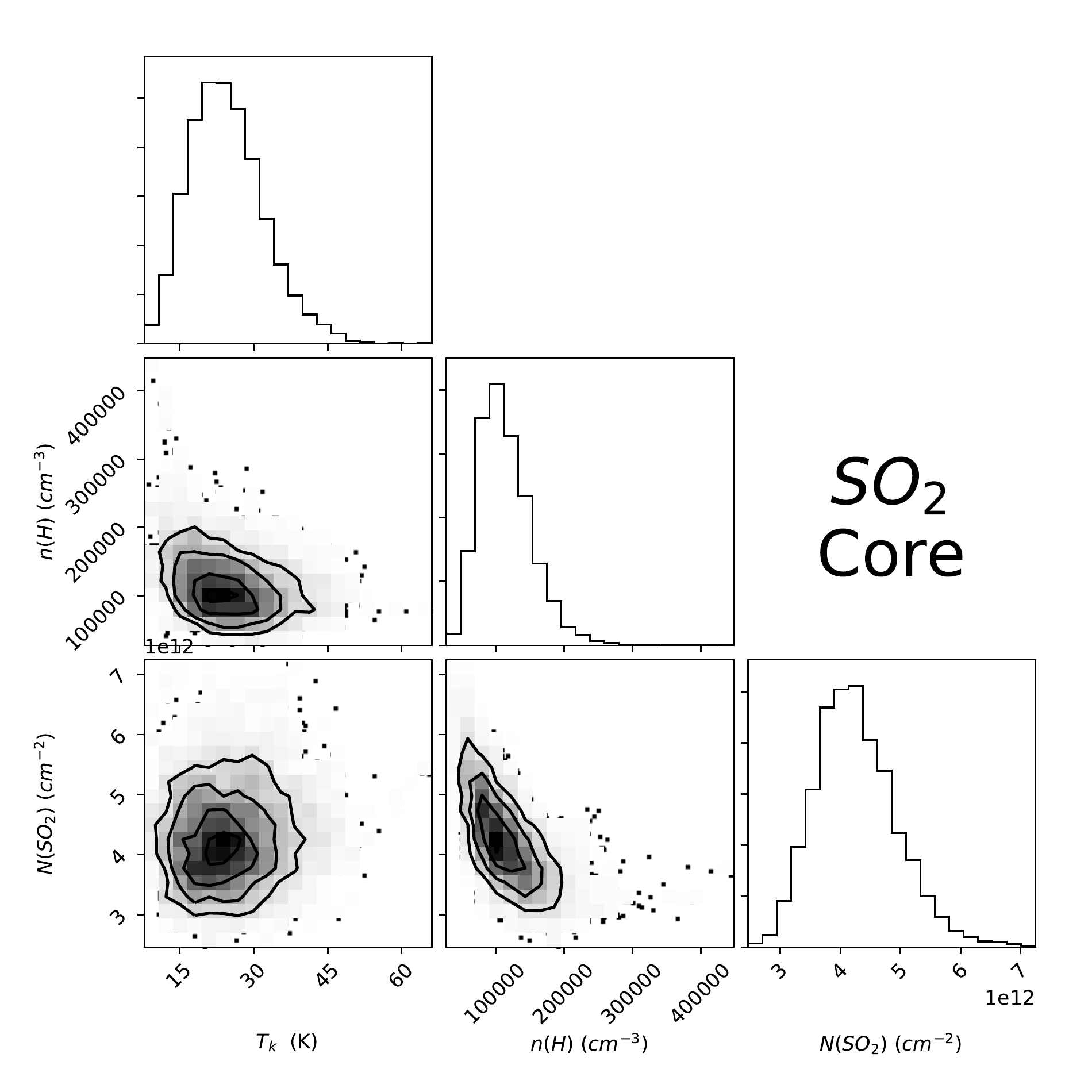}\\
 \includegraphics[trim=6mm 6mm 6mm 8mm, clip, scale=0.5]{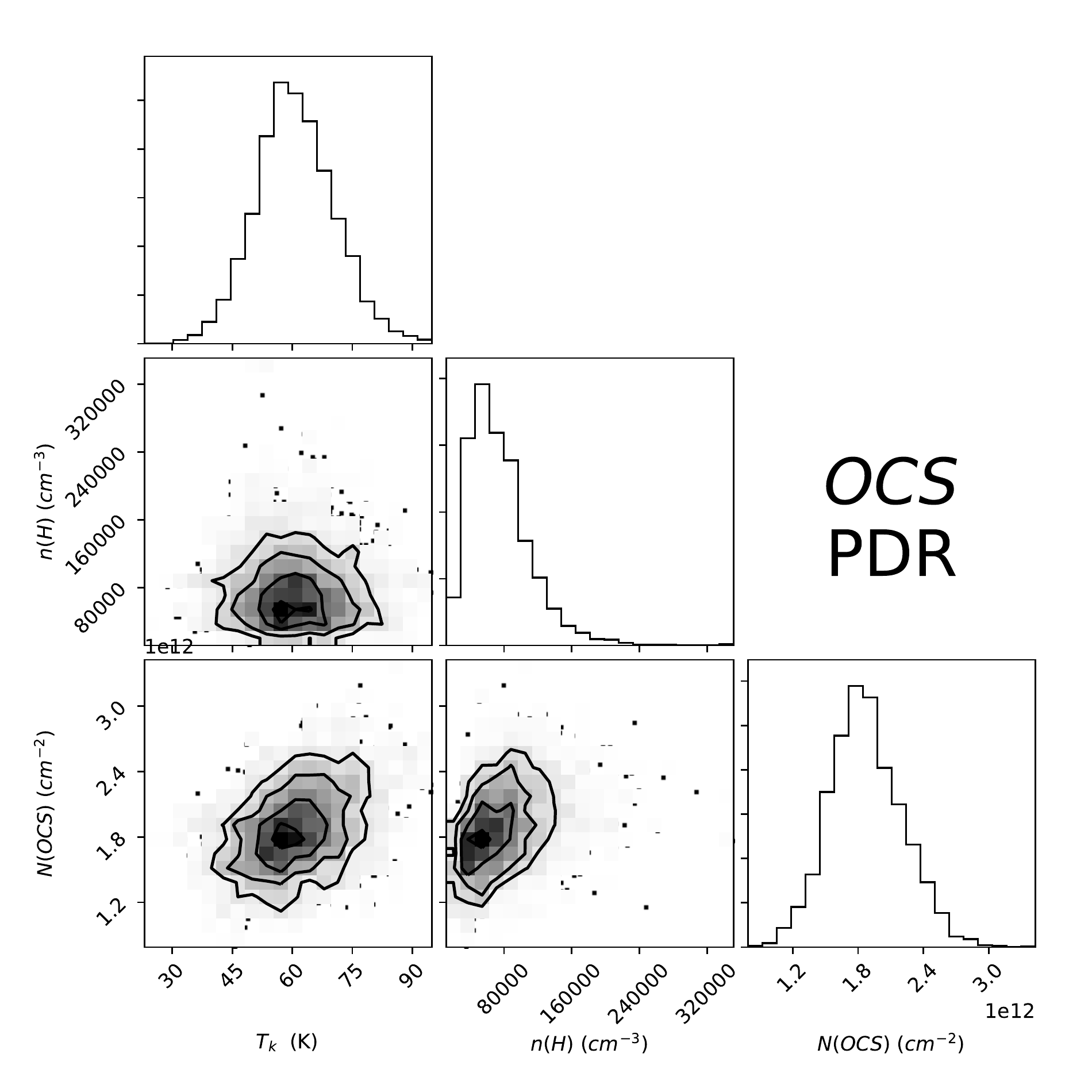}\includegraphics[trim=6mm 6mm 0mm 8mm, clip, scale=0.5]{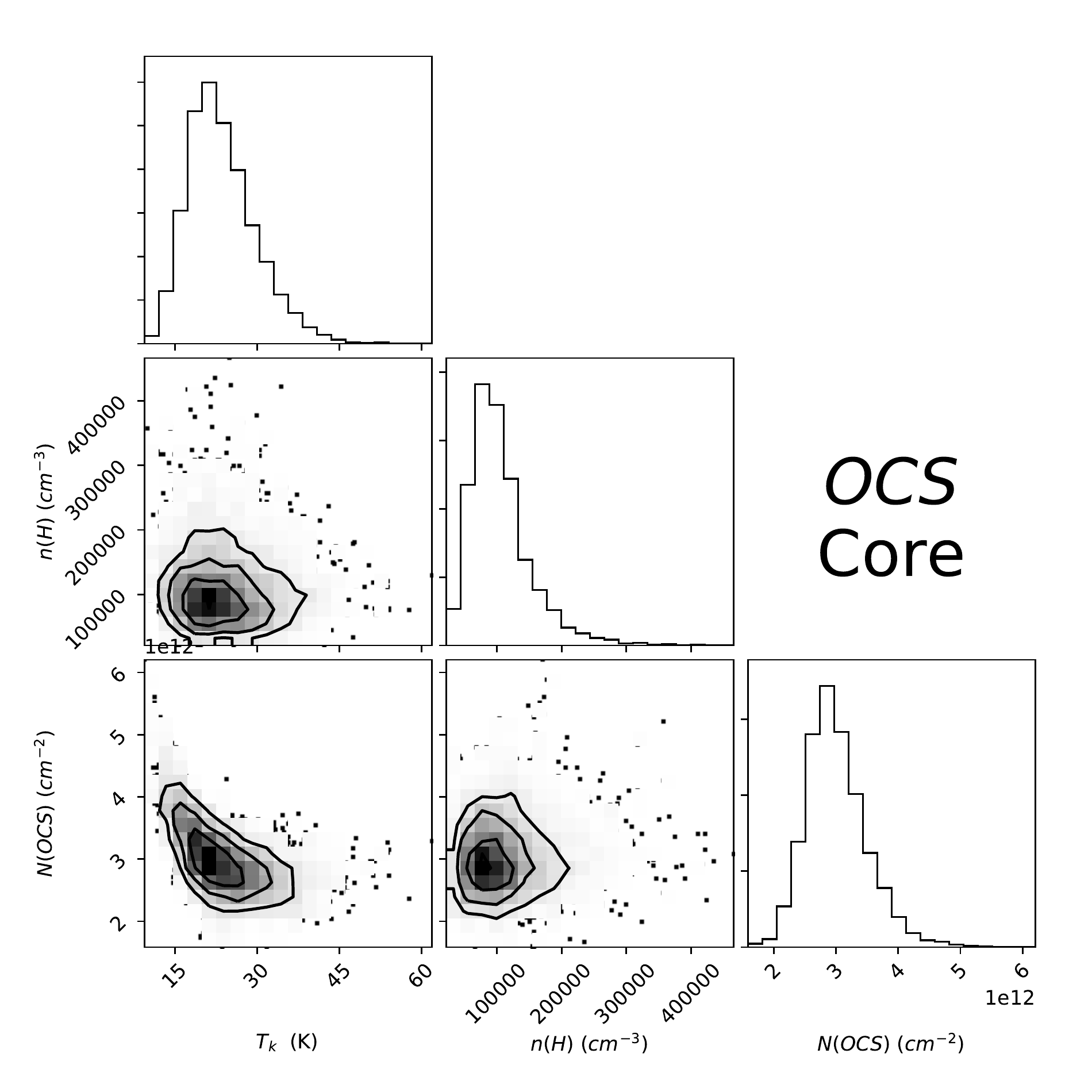}\\ 
\caption{Histograms of the MCMC RADEX parameter distribution for the different molecules and regions studied.}
\label{Fig:MCMC_LVG}
\end{center}
\end{figure*}

\addtocounter{figure}{-1}
\begin{figure*}[!h]
\begin{center}
 \includegraphics[trim=6mm 6mm 6mm 8mm, clip, scale=0.5]{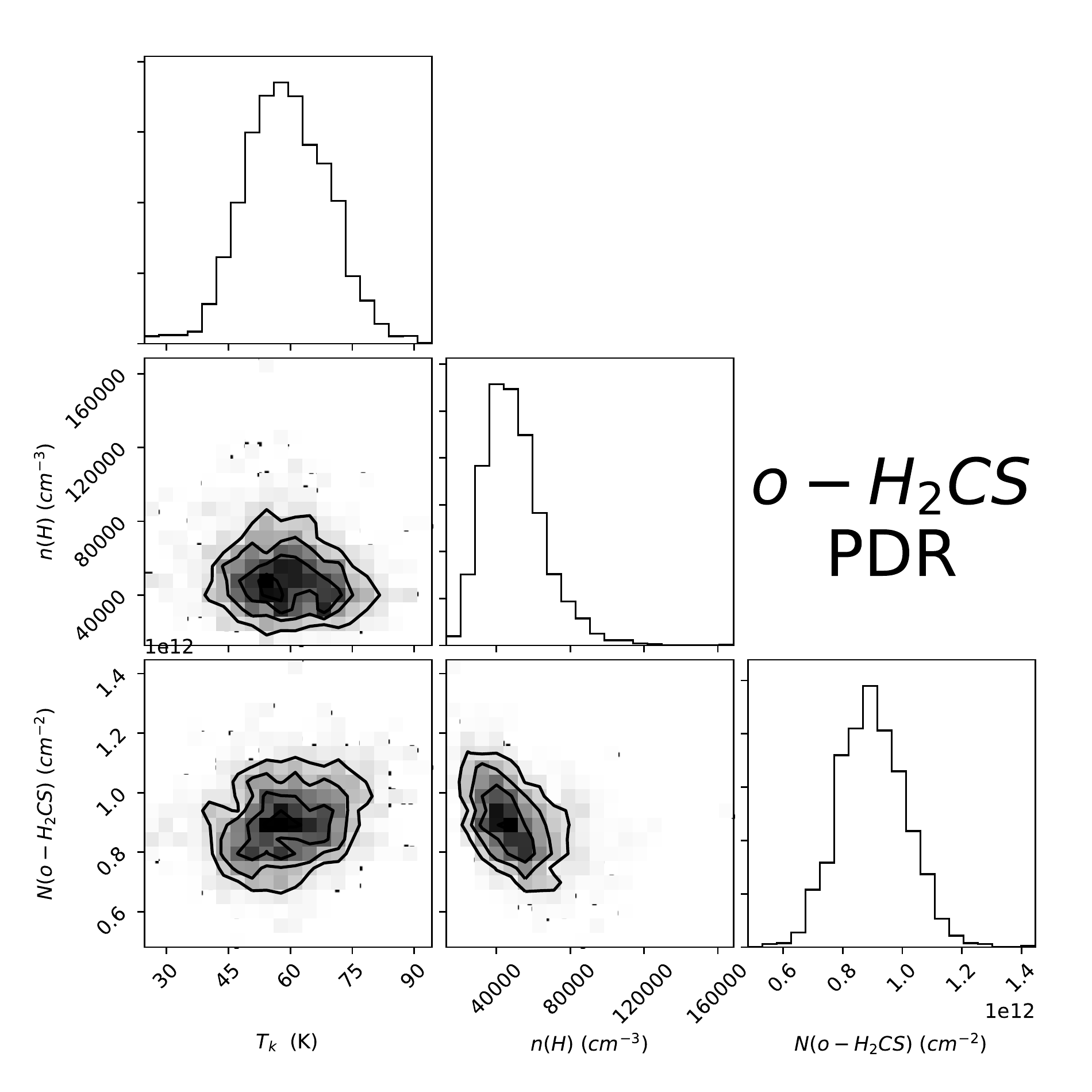}\includegraphics[trim=6mm 6mm 0mm 8mm, clip, scale=0.5]{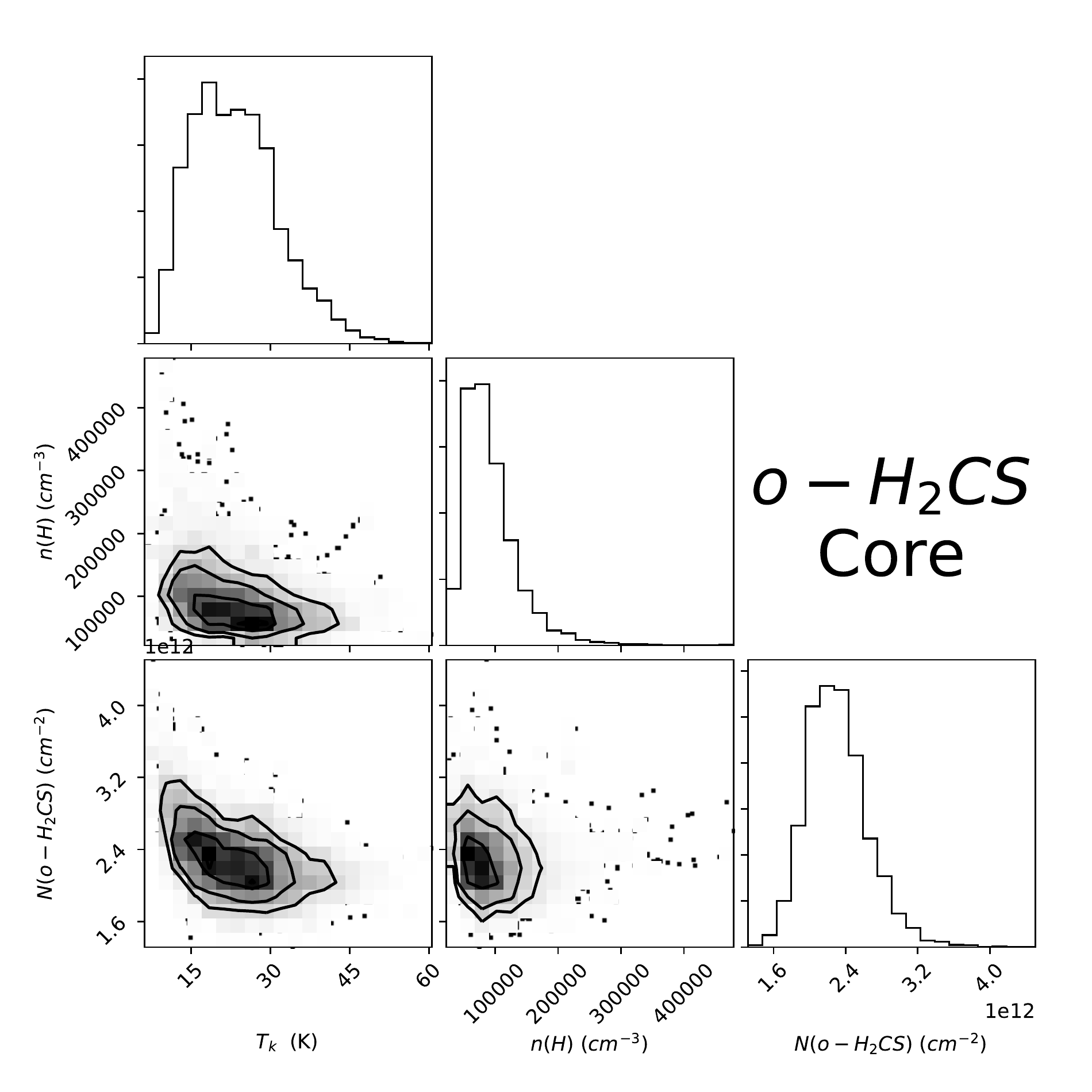}\\
 \includegraphics[trim=6mm 6mm 6mm 8mm, clip, scale=0.5]{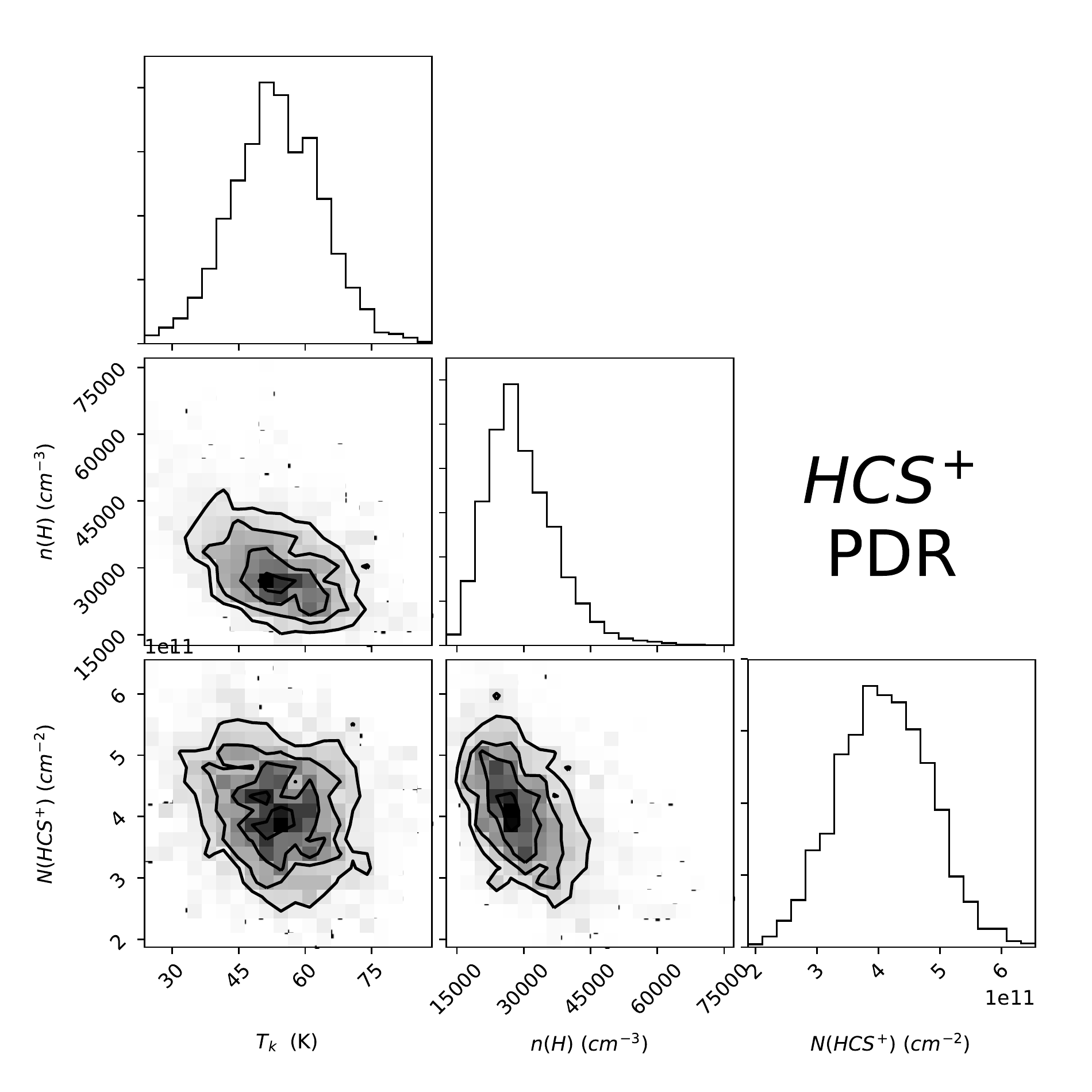}\includegraphics[trim=6mm 6mm 0mm 8mm, clip, scale=0.5]{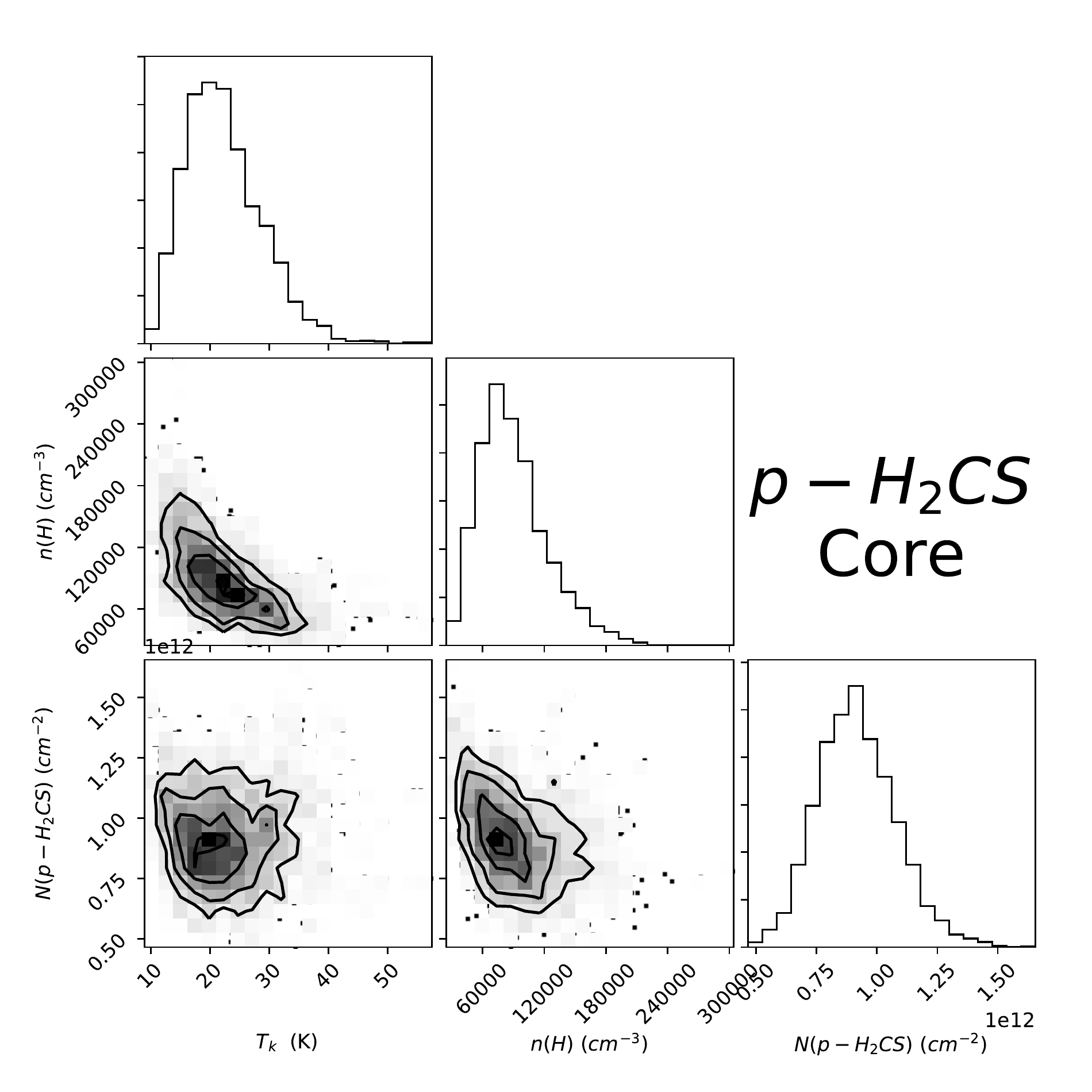}\\  
\caption{Histograms of the MCMC RADEX parameter distribution for the different molecules and regions studied.}
\label{Fig:MCMC_LVG}
\end{center}
\end{figure*}

\end{document}